%% file: main_NIPS.tex
\pdfoutput=1
\documentclass{article}

\usepackage[pdftex]{graphicx}
\usepackage[final]{neurips_2022}
\usepackage{wrapfig}
\usepackage[colorlinks,citecolor=gray]{hyperref} 

\usepackage[utf8]{inputenc} 
\usepackage[T1]{fontenc}    
\usepackage{url}            
\usepackage{booktabs}       
\usepackage{amsfonts}       
\usepackage{nicefrac}       
\usepackage{microtype}      
\usepackage{xcolor}         

\usepackage{subfigure}
\usepackage{amsmath}
\usepackage{amssymb}
\usepackage{xspace}
\usepackage{multirow}
\usepackage{stackengine}
\usepackage{wrapfig}
\usepackage{float}
\usepackage{color,soul}
\usepackage[hang,flushmargin]{footmisc} 

\newcommand{\myparagraph}[1]{\vspace{-5pt}\paragraph{#1}}
\input{macros}

\input{math_commands}

\title{Learning Neural Acoustic Fields}

\author{%
  Andrew Luo\\
  Carnegie Mellon University
  \And 
  Yilun Du\\
  Massachusetts Institute of Technology
  \And  
  {Michael J. Tarr} \\
  Carnegie Mellon University
  \And 
    Joshua B. Tenenbaum\\
   MIT BCS, CBMM, CSAIL
  \And 
  Antonio Torralba\\
  Massachusetts Institute of Technology
  \And 
  Chuang Gan\\
  UMass Amherst and MIT-IBM Watson AI Lab \\
}

\begin{document}

\maketitle
\input{text/00_abstract}
\input{text/01_intro}

\input{text/02_related_work}
\input{text/03_model}
\input{text/04_experiment}
\input{text/05_future}

\input{text/06_conclusion}
\input{text/09_acknowledgements.tex}

\newpage

\bibliography{ref}
\newpage


\input{appendix.tex}
\end{document}

%% file: macros.tex
\newcommand{\model}{NAFs\xspace}
\newcommand{\sect}[1]{Section~\ref{#1}}

\newcommand{\eqn}[1]{Eqn~(\ref{#1})}
\newcommand{\fig}[1]{Figure~\ref{#1}}

%% file: math_commands.tex
\usepackage{amsmath,amsfonts,bm}

\def\eqref#1{equation~\ref{#1}}

\def\1{\bm{1}}

\def\vq{{\bm{q}}}

\def\vv{{\bm{v}}}

\def\vx{{\bm{x}}}

\DeclareMathAlphabet{\mathsfit}{\encodingdefault}{\sfdefault}{m}{sl}
\SetMathAlphabet{\mathsfit}{bold}{\encodingdefault}{\sfdefault}{bx}{n}



%% file: text/00_abstract.tex
\begin{abstract}
Our environment is filled with rich and dynamic acoustic information. When we walk into a cathedral, the reverberations as much as appearance inform us of the sanctuary's wide open space. Similarly, as an object moves around us, we expect the sound emitted to also exhibit this movement. While recent advances in learned implicit functions have led to increasingly higher quality representations of the visual world, there have not been commensurate advances in learning spatial auditory representations. To address this gap, we introduce Neural Acoustic Fields (NAFs), an implicit representation that captures how sounds propagate in a physical scene. By modeling acoustic propagation in a scene as a linear time-invariant system, NAFs learn to continuously map all emitter and listener location pairs to a neural impulse response function that can then be applied to arbitrary sounds. We demonstrate NAFs on both synthetic and real data, and show that the continuous nature of NAFs enables us to render spatial acoustics for a listener at arbitrary locations. We further show that the representation learned by NAFs can help improve visual learning with sparse views. Finally we show that a representation informative of scene structure emerges during the learning of NAFs. Project site: \href{https://www.andrew.cmu.edu/user/afluo/Neural_Acoustic_Fields}{https://www.andrew.cmu.edu/user/afluo/Neural\_Acoustic\_Fields}
\end{abstract} 

%% file: text/01_intro.tex
\section{Introduction}
\label{sec:intro}

The sound of the ball leaving the bat, as much as its visible trajectory, tells us whether the hit is likely to be a home run or not. Our experience of the world around us is rich and multimodal, depending on integrated input from multiple sensory modalities. In particular, spatial acoustic cues provide us with a sense of the direction and distance of a sound source without needing visual confirmation, allow us to estimate the properties of a surrounding environment, and are critical to subjective realism in gaming and virtual simulations.

Recent progress in implicit neural representations has enabled the construction of continuous, differentiable representations of the visual world directly from raw image observations \citep{sitzmann2019srns,mildenhall2020nerf,Niemeyer2020DVR, yariv2020multiview}. However, our perception of the physical world is informed not only by our visual observations, but also by the spatial acoustic cues present in the environment. As a preliminary step in learning the acoustic properties of scenes, we explore an implicit model that represents the underlying impulse response of audio reverberations. As shown in \fig{fig:teaser}, our model can model the spatial propagation of sound in a physical scene.

Past work has explored capturing the underlying acoustics of a scene \citep{raghuvanshi2014parametric, raghuvanshi2018parametric, chaitanya2020directional}. These models, however, require handcrafted parameterizations which, critically, prevent such approaches from being applied to arbitrary scenes. In this work, we extend this approach by constructing an implicit neural representation which captures, in a \textit{generic manner}, the underlying acoustics of a scene. 

Learning a representation of scene acoustics poses several challenges compared to the visual setting. First, how do we generate plausible audio impulse responses at each emitter-listener position? While we may represent the visual appearance of a scene with an underlying three-dimensional vector, an acoustic reverberation (represented as an impulse response) can consist of over 10,000 time-domain values and, thus, is significantly harder to capture. Second, how do we learn an acoustic neural representation that densely generalizes to novel emitter-listener locations? In the visual setting, ray-tracing can enforce view consistency across large portions of a visual scene (modulo occlusions). While in principle, in a similar manner, we may reflect acoustic "rays" in a scene represented as an implicit function to obtain an impulse response, an intractable amount of compute is necessary to obtain the desired representation~\citep{nerv2021}.

To address both challenges, we propose Neural Acoustic Fields (\model). To capture the complex signal representation of impulse responses in a compact and spati ally continuous fashion, \model encode and represent an impulse-response in the time-frequency domain. Motivated by the strong influence of nearby geometry on anisotropic reflections~\citep{raghuvanshi2018parametric}, we propose to condition \model on local geometric information present at both the listener and emitter locations when decoding the impulse response. In our framework, local geometric information is learned directly from impulse responses. Such a decomposition facilitates the transfer of local information captured from training emitter-listener pairs to novel combinations of emitters and listeners.
\input{figtext/01_teaser}

We show that \model are able to outperform baselines in modeling scene acoustics, and provide detailed analysis of the design choices in \model. We further illustrate how the structure learned by \model can improve cross-modal generation of novel visual views of a scene. Finally, we illustrate how the learned representation of \model enable the downstream application of inferring scene structure.

%% file: figtext/01_teaser.tex
\begin{figure}[t!]
  \centering
\includegraphics[width=1.0\linewidth]{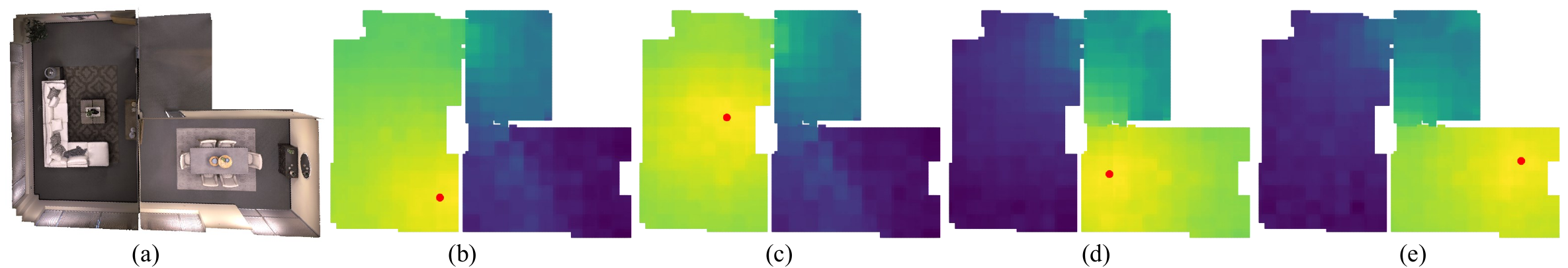}
   \vspace{-3mm}
  \caption{\small Neural Acoustic Field (NAF) learns an implicit representation for acoustic propagation. \textbf{(a)} A 3D top-down view of the house with two rooms. \textbf{(b)-(e)} The loudness of acoustic field as predicted by our NAF is visualized for an emitter located at the red dot. Notice how sound does not leak through walls, and the portaling effect open doorways can have. Louder regions are shown in yellow.}
  \vspace{-1.5em}
  \label{fig:teaser}
\end{figure}

%% file: text/02_related_work.tex
\section{Related Work}
\label{sec:related}
\myparagraph{Audio Field Coding}
There is a rich history of sound field representation, encoding, and interpolation methods for 3D spatial audio. Some approaches seek to directly approximate the sound field~\citep{mignot2013low,antonello2017room,ueno2018kernel} while adding handcrafted priors. Others adopt a parametric representation that seek to model only the perceptual cues~\citep{raghuvanshi2014parametric, raghuvanshi2018parametric, chaitanya2020directional, mehra2014source, ratnarajah2021fast}. Since the complete acoustic field of a scene is computationally prohibitive to simulate in real time, and expensive to store in full fidelity, these methods have typically relied on a handcrafted encoding of the acoustic field, prioritizing efficiency above reproduction fidelity. In recent years, there has been interest in using deep learning to directly learn a sound field from data, without making strong assumptions about the scene. However in practice, these approaches use either a stationary listener or emitter~\citep{richard2020neural,richard2022deep}. In contrast, our work enables the querying of the sound field for arbitrary emitter and listener locations.

\textbf{Implicit representations}
Our approach towards modeling the underlying acoustics a scene relies on the use of a neural implicit representations. Implicit representations have emerged as a promising representation of 3D geometry ~\citep{Niemeyer2019ICCV,chen2019learning,park2019deepsdf,saito2019pifu,hong20223d}  and appearance~\citep{sitzmann2019srns,mildenhall2020nerf,Niemeyer2020DVR,yariv2020multiview, wang2021ibrnet} of a scene. Compared to traditional discrete representations, implicit representations are a continuous mapping capable of capturing data at an "infinite resolution". In~\citep{jiang2020local} proposed a grid based representation for implicit scenes, while more recently~\citep{devries2021unconstrained} has adopted spatial conditioning for 3D image synthesis, where in both settings, the grid enables a higher-fidelity encoding of the scene. Our work also leverages local grids to model acoustics, but as an inductive bias and way to generalize to novel inputs. 

\input{figtext/arch}
\textbf{Audio-Visual Learning}
Our work is also closely related to joint modeling of vision and audio. By leveraging the correspondence between vision and audio, work has been done to learn unsupervised video and audio representations \citep{aytar2016soundnet, arandjelovic2017look}, localize objects that emit sound~\citep{senocak2018learning, zhao2018sound}, and jointly use vision and audio for navigation~\citep{chen2020soundspaces}. Recent work aims to propose plausible reverberations or sounds from image input~\citep{singh2021image2reverb, du2021gem}, these approaches model the STFT using either convolution or implicit functions, which we also utilize. Different from them, our work leverages the geometric features learned by modeling acoustic fields to improve the learning of 3D view generation. 

%% file: figtext/arch.tex
\begin{figure*}[!t]
  \centering
  \includegraphics[width=1.0\linewidth]{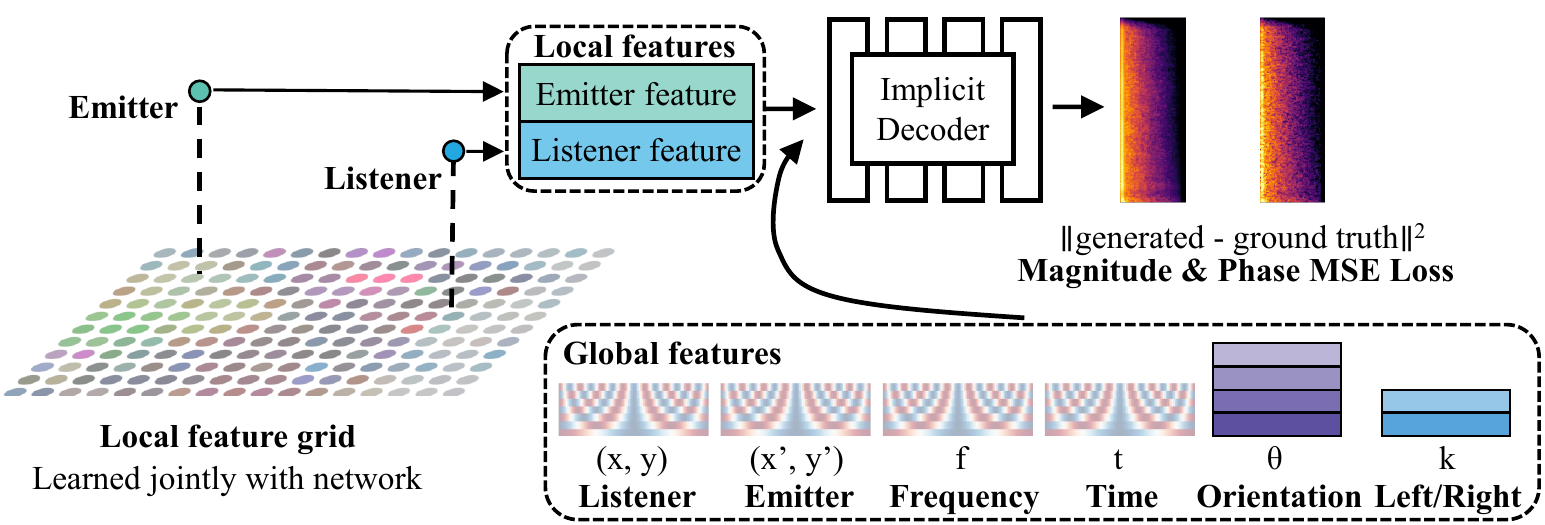}
  \caption{\small Overview of our NAF architecture where listener and emitter share a feature grid. Given a listener position and an emitter location, we first query a grid for local features which are learned together with the network during training. We compute the sinusoidal embedding of the positions, frequency, and time, and query a discrete embedding matrix using the orientation and left/right ear. Our method predicts  magnitude and phase.}
  \label{fig:netarch}
\end{figure*} 

%% file: text/03_model.tex
\section{Methods}
\vspace{-2.2mm}
\label{sec:methods}
We are interested in learning a generic acoustic representation of an arbitrary scene, which can capture the underlying sound propagation of arbitrary sound sources across both seen and unseen locations in a scene.
We first review relevant background information towards modeling environment reverberations. We then describe Neural Acoustic Fields (NAFs), a neural field which we show can capture, in a generic manner, the acoustics of arbitrary scenes. We further discuss how we can parameterize NAF so that it can capture acoustics property even at unseen sound sources and listener positions. Finally, we discuss the implementation details of our model illustrated in Figure~\ref{fig:netarch}. 

\subsection{Background on the Propagation of Sound}
\label{sect:background}
The sound emitted by a sound source undergoes decay, occlusion, and scattering due to both the geometric and material properties of a scene. For a fixed location pair $(\vq, \vq')$, we define the impulse-response at a listener position $\vq$, as the sound pressure $p(t; \vq, \vq')$ induced by an impulse at $\vq'$.

Given an accurate model of the impulse-response $p(t; \vq, \vq')$, we may model audio reverberation of any sound waveform  $s(t)$ emitted at $\vq'$, by computing the response $r(t, \vq, \vq')$ at listener location $\vq$ by querying the continuous field and using temporal convolution:
\begin{align}
    r(t; \vq, \vq') = s(t) \circledast p(t; \vq, \vq')
    \label{eqn:convolve}
\end{align}

\subsection{Neural Acoustic Fields}
\label{sect:naf}
We are interested in constructing a continuous representation of the underlying acoustics of a scene, which may specify the reverberation patterns of an arbitrary sound source. The parameterization of an impulse-response introduced in \sect{sect:background} provides us with a method to model audio propagation when given an omnidirectional listener and emitter. To construct a model of a directional listener, we need to further model the 3D head orientation $\theta \in \mathbb{R}^2$, and ear $k \in \{0,1\}$ (binary left or right) of a listener, in addition to the spatial position $\vq \in \mathbb{R}^3$ of the listener and $\vq' \in \mathbb{R}^3$ of the emitter.

We may then model the time domain impulse response $\vv$ using a neural field $\Phi$ which takes as input the listener and emitter parameters:
\begin{align}
    \Phi : \mathbb{R}^{8} \times \{0,1\} \rightarrow \mathbb{R}^T ; (\vq, \theta, k, \vq') \rightarrow \Phi(\vq, \theta, k, \vq') = \vv
\end{align}



Directly outputting the impulse-response waveform $\vv \in \mathbb{R}^T$ in the time domain with a neural network is difficult due to its high dimensional (over 10,000 elements) and chaotic nature. A na\"ive solution would be to further add $t$ as an additional argument to our neural field, but we found that such a solution worked poorly, due to the highly non-smooth representation of the waveform (see supplementary). We instead encode the impulse-response utilizing a short-time Fourier transform (STFT) denoted $v_\text{STFT}$, which we find to be significantly more amenable to neural network prediction due to the smoother nature of the time-frequency space. In Figure~\ref{fig:impulse} we show magnitude spectrograms for ground truth impulse responses and those learned by our network.
\input{figtext/impulses}
As $v_\text{STFT}$ is a complex value, we further factorize $v_\text{STFT}$ into log-magnitude and phase angle components. For phase angle, we use the instantaneous frequency (IF) representation proposed in GANSynth~\citep{engel2019gansynth}. To compute the IF representation, the phase angle is unwrapped and has the finite difference taken across the time dimension for each frequency in the STFT. This transformation results in a phase representation that conducive to learning due to more regular structure.

Thus, our parameterization of NAF is a neural field $\Omega$ that is trained to estimate the impulse response function $\phi$, and outputs $[\vv_{\text{STFT\_mag}}, \vv_{\text{STFT\_IF}}]$ for a given time and frequency coordinate:
\begin{align}
    & \Omega :  \mathbb{R}^{10} \times \{0,1\} \rightarrow \mathbb{C} \nonumber \\
    & (\vq, \theta, k, \vq', t, f) \rightarrow \Omega(\vq, \theta, k, \vq', t, f) \approx [\vv_\text{STFT\_mag}(t,f), \vv_\text{STFT\_IF}(t,f)]
    \label{eqn:naf_freq}
\end{align}
We train our model using MSE loss between the generated and ground truth spectrograms $\vv_\text{STFT}$:
\begin{align}
\mathcal{L}_{\text{NAF}} =&  \| \Omega(\vq, \theta, k, \vq', t, f)_\text{mag} - \vv_\text{STFT\_mag}(t, f) \|^2 \nonumber +\\ \alpha &\| \Omega(\vq, \theta, k, \vq', t, f)_\text{IF} - \vv_\text{STFT\_IF}(t, f) \|^2
\label{nafeqn}
\end{align}
across spectrogram coordinates $t$ and $f$. Where $\alpha$ is a scaling value used to balance the two losses.
\subsection{Generalization through Local Geometric Conditioning}
We are interested in parameterizing the underlying acoustic field, so that we may not only accurately represent impulse-response at emitter-listener pairs we see during training, but also at novel combinations of emitter and listener seen at test time. Such generalization may be problematic when directly parameterizing \model utilizing a MLP with inputs specified in \eqn{eqn:naf_freq}, as the network may learn to directly overfit and entangle the relation between emitter and listener impulse-responses.

What generic information may we extract from a given impulse-response between an emitter and listener? In principle, extracting the full dense geometric information in a scene would enable us to robustly generalize to new emitter and listener locations. However, the amount of geometric information available in a particular impulse-response, especially for positions far away from either current emitter and listener is limited, since these positions have little impact on the underlying impulse-response. In contrast, the local geometry near either emitter and listener positions will have a strong influence in the impulse-response, as much of the anisotropic reflection comes from such geometry~\citep{paasonen2017proximity}. Inspired by this observation, we aim to capture and utilize local geometric information, near either emitter or listener locations, as a means to predict impulse-responses across novel combinations. 

To parameterize and represent these local geometric features, we learn a 2D grid of spatial latents which we illustrate in \fig{fig:netarch}. The spatial latents are randomly initialized and uniformly distributed in the room. When predicting an impulse-response at a given emitter and offset position, we query the learned grid features at both emitter and listener positions, and provide it as additional context into our NAF network $\Omega$. Such features provide rich information on the impulse-response, enabling NAF to generalize better to unseen combinations of both emitter and listener locations. In the rest of this work, we refer to the \model with local geometric features as $\Omega_{grid}$. We learn grid latent features jointly with the underlying parameters of NAF. Additional details can be found in the supplementary.

Such a design choice, however, still requires us to consider how to further combine local geometric information captured separately from either listeners or emitters. A na\"ive implementation would be to maintain separate feature grids for both listener and emitter positions. Such an approach fails to account for the fact that the local geometric information captured by emitter may also inform the local geometric information around a listener. Examining Green's function, which is the solution to the wave equation, we note that it is in fact symmetric with respect to exchanging either listener or emitter positions~\citep{chaitanya2020directional}, indicating that the impulse-response does not change when omnidirectional emitters and listeners are swapped (acoustic reciprocity). Such a result means that we may in fact utilize the local geometric information captured near an emitter position interchangeably for either emitters and listeners. Thus, we propose our local geometric information as a single latent grid, which we show to outperform the na\"ive dual grid implementation.

%% file: figtext/impulses.tex
\begin{figure*}[!t]
  \centering
  \includegraphics[width=1.0\linewidth]{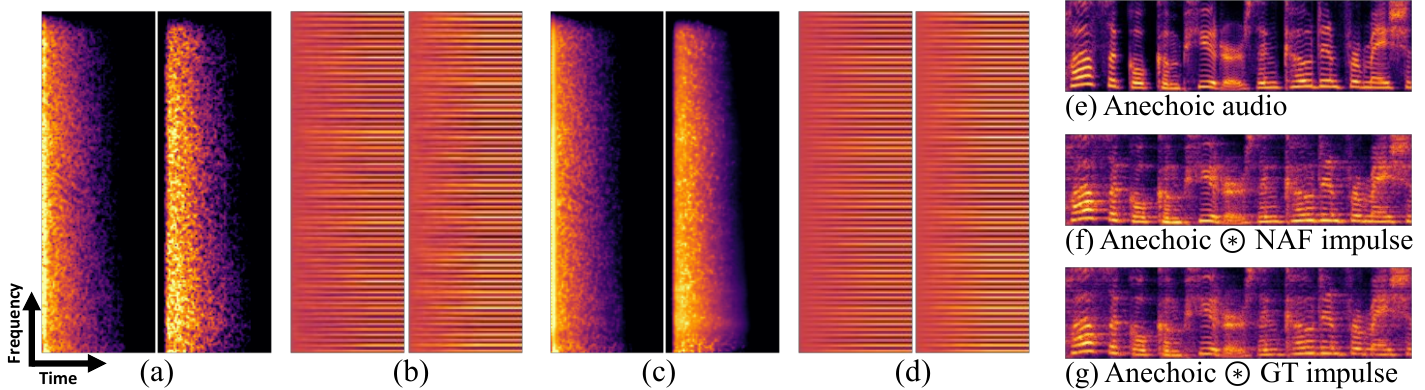}
  \caption{\small \textbf{Qualitative Visualization of Test Set Impulse Response Prediction.} \textbf{(a)} Ground truth log-magnitude. \textbf{(b)} Ground truth phase. \textbf{(c)} NAF predicted log-magnitude. \textbf{(d)} NAF predicted phase. \textbf{(e)} log-magnitude of an anechoic audio (without any reverberation). \textbf{(d)} The sound with a reverberation impulse response from our NAF. \textbf{(e)} The sound with the ground truth reverberation impulse response applied. Note phases are unwrapped for visualization purposes. Time \& frequency are on the horizontal and vertical axes.}
  \label{fig:impulse}
\end{figure*}

%% file: text/04_experiment.tex
\section{Experiments}\label{sec:exp}
\input{figtext/main}
In this section, we demonstrate that our model can faithfully represent the acoustic impulse response at seen and unseen locations. Additional ablation studies verify the importance of utilizing local geometric features to enable test time generation fidelity. Next, we demonstrate that learning acoustic fields could facilitate improved visual representations when training images are sparse. Finally we show that the learned NAF can be used to infer scene structure. 

\subsection{Setup}
For each scene, we holdout $10\%$ of the RIRs randomly as a test set. Each scene is trained for $200$ epochs, which takes around $6$ hours for the largest scenes on four \texttt{Nvidia V100s}. In each batch, we sample $20$ impulse responses, and randomly select $2,000$ frequency \& time pairs within each spectrogram. An initial learning rate of $5\times10^{-4}$ is used for the network and the grid features. We add a small amount of noise sampled from $ \mathcal{N}(0, 0.1)$ to each coordinate during training to prevent degenerate solutions. For evaluating the learned acoustic fields, we use two different datasets:\myparagraph{Soundspaces.} Soundspaces~\citep{chen2020soundspaces, straub2019replica} is a synthetic dataset generated via ray-tracing, and is auralized with an ambisonic head-related transfer function (HRTF). This dataset consists of $R_i$ probe points for each scene, with each probe capable of representing an emitter or listener location for up to $R_{i}^{2}$ emitter and listener pairs. The emitters are represented as omnidirectional, while the listener acts as a stereo receiver that can have one of four different orientations. The listeners and emitters are at fixed height. Our NAFs are trained on $6$ representative scenes, where $2$ consist of multi-room layouts; $2$ consist of a single room with a non-rectangular walls; and $2$ consist of a single room with rectangular walls as in Figure~\ref{fig:main}.
\myparagraph{MeshRIR.} The MeshRIR dataset~\citep{koyama2021meshrir} is recorded from a real scene, and contains monaural data collected from a cuboidal room. The listener locations are at fixed height. The emitters surround the listeners both above and below the listener plane.

\subsection{Architecture Details}
The Soundspaces dataset lacks the full parameterization of an acoustic field described in  Equation~\ref{eqn:naf_freq}, so we train NAF with a restricted parameterization that is available in the dataset. This allows for two degrees of freedom along the $x-y$ plane for the listener locations $q \in \mathbb{R}^2$ and the emitter location $q' \in \mathbb{R}^2$. The listener is binaural with $k \in \{0,1\}$, and can assume four possible orientations $\theta \in \{0, 90, 180, 270\}$, while the emitter is omnidirectional.  In particular, we utilize a parameterization of $\Omega_{\text{grid}}$ which maps an input tuple $[x, y, x', y',f, t] \times{\{0, 90, 180, 270\}}\times{\{0, 1\}}$ to two scalar values that represents the magnitude and phase for a given time and frequency in the STFT:
\begin{align}
    \Omega_{\text{grid}}(x, y, \theta, k, x', y', t, f) \Rightarrow  [\vv_{\text{STFT\_mag}}(t,f), \vv_{\text{STFT\_IF}}(t,f)]
\end{align}
To encode the rotation $\theta$, as there are only 4 possible discrete rotations in the dataset, we directly query into a learnable embedding matrix of shape $\mathbb{R}^{4 \times n}$, returning a $\mathbb{R}^{1 \times n}$ vector. Similarily, to encode the left and right ear $k$, we similarly query into a learnable embedding matrix of shape $\mathbb{R}^{2 \times n}$, returning a $\mathbb{R}^{1 \times n}$ vector. The $f, t$ tuple representing the frequency and time respectively are scaled to $(-1, 1)$ and processed with sinusoidal encoding using 10 frequencies of $\sin$ and $\cos$. For MeshRIR, we set emitter $q' \in \mathbb{R}^3$ to account for the emitters that can vary in height, and do not utilize the orientation or binaural embedding.
\input{figtext/full_comp}

To obtain local geometric features for either an emitter or listener in a scene, we assume that our scene is contained within a set of pixels $\mathcal{P} = \{P_1 ... P_k\}$ which form a grid over the scene. For a given position tuple $(x, y)$ as query location, we then interpolate the local features. Where $\mathcal{L}(\cdot)$ is the interpolation function. $(p^{*}_1 \dotsc p^{*}_k)$ are the set of all pixel that form the grid, and $\tilde{f}(\cdot)$ represents the features stored at a given pixel:

\begin{align}
(x, y) &\Rightarrow \mathcal{L}(x,y; \tilde{f}(p^{*}_1), \dotsc \tilde{f}(p^{*}_k))\\
 &=\sum_{i=1}^{k} w_{i}\tilde{f}(p^{*}_{i})
\end{align}
 $w_i$ is determined by a Nadaraya-Watson estimator with a Gaussian weighting kernel applied to the distance between query and grid coordinates:
 \begin{align}
    &w_i = K((x,y), (x_i, y_i))/\sum_{j=1}^{k} K((x,y), (x_j, y_j))\\
&K(\vx, \vx') = \exp(-\|\vx-\vx'\|^{2}_{2}/2\sigma^{2})
\end{align}
  Because this interpolation function is differentiable, we jointly learn the grid features during training. These queried features are combined with the coordinates processed with sinusoidal encoding using 10 frequencies of $\sin$ and $\cos$ functions. We process both the listener and emitter position tuples this way. We combine the grid based features with the sinusoidal embeddings and the discrete indexed embeddings as the input to our multilayer perceptron $f_\phi$. Please refer to Figure~\ref{fig:netarch} for a visualization of our model, and supplementary for further details. We compare using a shared local geometric feature with the emitter and listener, as well as having the emitter and listener query their own individual grids. 

\subsection{Evaluation of Neural Acoustic Fields}
We first validate that we can capture environmental acoustics at unseen emitter-listener positions. 

\input{figtext/storage_maintxt}
\noindent\textbf{Baselines.} We compare our model against two widely used high performance audio coding methods: Advanced Audio Coding (AAC) and Xiph Opus. We use low bitrates in order to attempt to approach the storage costs for our NAF. For each method, we apply both linear and nearest neighbor interpolation to the coded acoustic fields. Both linear and nearest neighbor approaches are widely used~\citep{savioja1999creating,raghuvanshi2010precomputed,porschmann2020comparison} in modeling of spatial audio. We further implement the binaural DSP baseline described by~\citep{richard2020neural}, which uses the image source method and a KEMAR HRTF. We also compare the listener and emitter either sharing or using individual local geometric features in our NAFs. 

Each method is provided with the same train-test split. We visualize the acoustic fields produced by different methods in \fig{fig:full}. Details of our baselines can be found in the supplementary section G.

\noindent\textbf{Metrics.}  We evaluate the results of our synthesis by measuring the spectral loss~\citep{defossez2018sing} between the generated and the ground truth log-spectrograms, as well as measuring the percentage error between the T60 reverberation time in the time domain. In this case, lower spectral loss and T60-error values indicates a better result. Additional quantitative results can be found in the supplementary.
We also perform a human evaluation where subjects are presented with a two-alternative forced-choice task. Each trial requires selecting if the NAFs or the AAC-nearest auralized music samples best match with ground truth auralization. 
\input{figtext/learning_NAFs_test}
\input{figtext/sparse_nerf_table}\\
\noindent\textbf{Results.} As shown in Table~\ref{table:results}, our NAFs achieve significantly higher quality on the modeling of unseen impulse responses compared to strong interpolation baselines across all six scenes. Observing the qualitative results in Figure~\ref{fig:main}, we observe that NAFs can predict smoothly varying acoustic fields that are affected by the physical surroundings. Extending our model to MeshRIR which is captured from a real scene, we observe that our NAFs continue to perform better on both spectral and T60 metrics.
Comparing our results against baselines in \fig{fig:full} and  Table~\ref{fig:storagetable}, our methods are able to better approximate the ground truth at a fraction of the storage cost, and does not exhibit the sound energy leakage present in linear interpolation. The size of a spatial acoustic field is important for real life applications. In our human evaluation with 21 subjects who were asked to judge 10 test-time RIRs, 82.38\% of responses indicates that our NAFs were higher quality compared to AAC-Nearest.

A comparison of using shared and dual local geometric features indicates that despite having fewer learnable parameters, we achieve better performance by sharing the local geometric features. Examples of individual impulse responses generated by our model are shown in Figure~\ref{fig:impulse}.

\noindent\textbf{Generalization through Geometric Conditioning.} We next assess the impact of utilizing local geometric conditioning as a means to generalize to novel combinations of emitter-listener positions. On the "Large 1" room,  in Figure~\ref{fig:ablation} we evaluate test set spectral error when NAF is trained with a limited percentage of the training data either with or without local geometric conditioning. We find that such geometric conditioning enables better test set reconstruction error, with the performance gap increasing with less data.
\begin{minipage}{\textwidth}
  \begin{minipage}[b]{0.54\textwidth}
 \input{figtext/sparse_nerf}
  \end{minipage}
  \hfill
  \begin{minipage}[b]{0.41\textwidth}
    \centering
    \input{figtext/ablation}
    \end{minipage}
  \end{minipage}

\subsection{Cross-modal learning}

In this experiment, we explore the effect of jointly learning acoustics and visual information when we are given sparse visual information. Recall that our NAF includes a local geometric feature grid $\mathcal{P}$ that covers the entire scene. For our cross-modal learning experiment, we jointly learn this feature grid with a NeRF network modified to accept local features sampled from this grid along with the traditional sinusoidal embedding. In the acoustics branch, we query the grid using emitter and listener positions. In the NeRF branch, we use point samples along the ray projected on the grid plane to query the features. In both cases, the process is fully differentiable. We use a standard implementation of NeRF with a coarse and fine network.

In the NeRF only setting, we minimize color $C$ reconstruction loss for a ray $r$ over a batch of rays $\mathcal{R}$: $\mathcal{L}_{\text{RGB}} = \sum_{r \in \mathcal{R}}||\hat{C}(r)-C(r)||^{2}_{2}$. In contrast, in the NAF + NeRF experiment, we jointly minimize $\mathcal{L}_{\text{RGB}} + \mathcal{L}_{\text{NAF}}$, where $\mathcal{L}_{\text{NAF}}$ is defined in equation~\ref{nafeqn}. We utilize 64 coarse samples and 128 fine samples for each ray, and sample 1024 rays per batch.

\noindent\textbf{Results.} We train on the two large rooms in our training set. For each room $75, 100, 150$ images are used for training, while the same $50$ images of novel views are used for all configurations during testing. In Table~\ref{fig:nerf} we observe that training with acoustic information helps improve the PSNR and SSIM of the visual output. This effect is more significant when the training images are very sparse, the NAF network helps less when there is sufficient visual information. Qualitative results are shown in Figure~\ref{fig:sparse}, we see there is a reduction of floaters in free space. 

\subsection{Inferring scene structure}

Given a reverberant sound, humans are able to build a mental representation of the surrounding room and make a judgement about the distance of nearby obstacles~\cite{kolarik2016auditory}. We investigate the intermediate representations constructed by our neural network in the process of learning an acoustic field, and examine if these representations can be used to decode the scene structure.

\noindent\textbf{Setup.} The intermediate representation of the NAF depends on both listener locations $q$ and emitter location $q'$, the rotation angle $\theta$, the ear $k$, the time $t$ and frequency $f$. For consistency, at a given location $(x^{*},y^{*})$ in the scene, we extract the NAF latent by setting the emitter location $q_{i}' = (x^{*},y^{*})_{i}$. For the listener location, we iterate over five randomly selected points in the scene $q \in [q_1, \dotsc, q_5]$, which we keep constant for all $q_{i}'$. The rotation angle is fixed to $\theta = 0$, and we compute the representation average over all possible $(k, t, f)$, and concatenate latents for the selected $q$. For our NAFs, latents are extracted from the last layer prior to the output which includes $512$ neurons.
\input{figtext/linear_probe_merged}
\input{figtext/lin_probe_table}
As a comparison to our learned representation, we extract Mel-frequency cepstral coefficients (MFCCs) from the ground truth impulse response provided by a nearest neighbor interpolator. We use a similar setup as above, for a given location we set this to be $q_{i}'$, and iterate over the same five listener locations $q_{1 \dotsc 5}$. We average the MFCCs over the left and right ear, and concatenate for the selected $q$. After flattening, the MFCC features are approximately 500 dimensional for any
given room.

We fit a single linear layer to NAF and MFCC features respectively. For testing and visualization of the linear decoding results, we sample a regular grid of points with $0.1m$ distance between each point. For fitting the linear decoder, we randomly sample points within the scene such that the number of training points are $10\%$ as many as the testing points. For each location in the scene, we extract the distance to the nearest wall as the decoding target. 

\noindent\textbf{Results.} We visualize the results of our linear decoding in \fig{fig:linearmap}. The intermediate representation of our \model reveals an underlying structure that is both smooth and semantically meaningful. In the multiroom scenes, the latent is well separated for each room. We are able to successfully decode the scene structure with a linear layer when using our \model, but decoding fails when using MFCC features. In Table~\ref{fig:linprobetable}, we show the amount of explained variance of our decoding results on the test set. Our learned features are able to consistently achieve much higher scores than those using MFCC features. 

%% file: figtext/main.tex
\begin{figure*}[!t]
  \centering
  \includegraphics[width=1.0\linewidth]{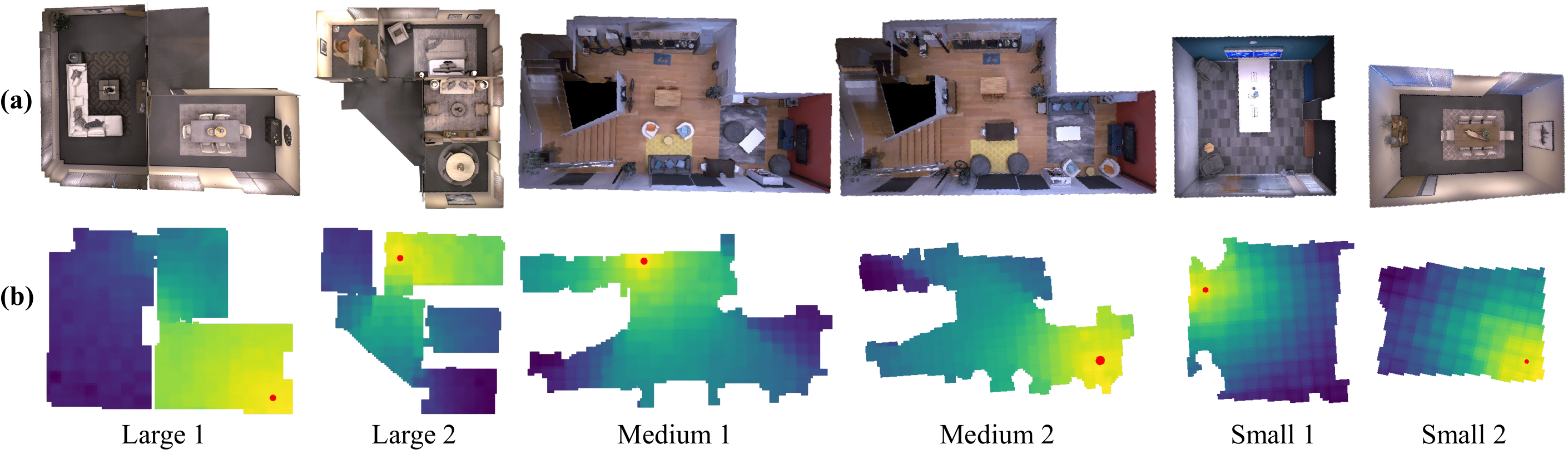}
  \caption{\small \textbf{Qualitative Visualization of Neural Acoustic Fields.} \textbf{(a)} Top down view of the rooms.  \textbf{(b)} Results as inferred by our neural acooustic field. Loudness of a sound given a emitter location indicated in red, lighter color indicates louder sound. Note how openings and walls lead to portaling and occlusion of the sound.}
  \label{fig:main}
\end{figure*}

%% file: figtext/full_comp.tex
\begin{figure*}[!t]
  \centering
  \includegraphics[width=1.0\linewidth]{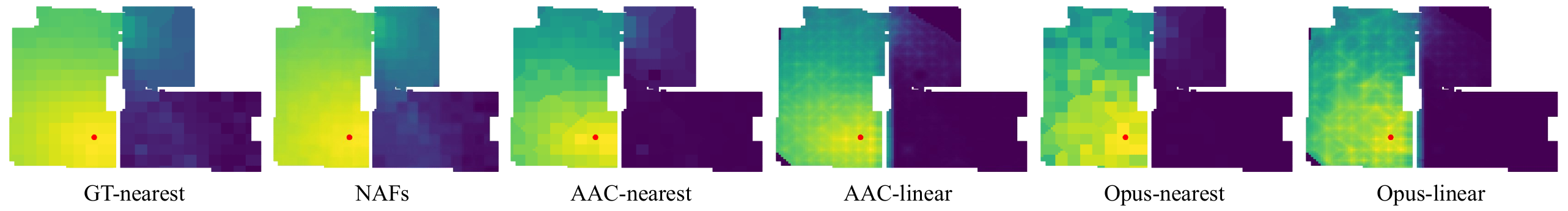}
  \caption{\small \textbf{Comparison of the acoustic fields.} From left to right, we visualize the loudness maps generated by the full ground truth, our NAFs, and by AAC or Opus coding combined with linear and nearest neighbor interpolation on the training set. Emitter location shown in red. Our method can faithfully reproduce the loudness map present in the ground truth.} 
  \label{fig:full}
\end{figure*}

%% file: figtext/storage_maintxt.tex


\begin{wraptable}{r}{4.5cm}
\setlength{\tabcolsep}{4.5pt}
\centering
\scalebox{0.8}{
\begin{tabular}{lcll}
\textbf{Method} & \multicolumn{3}{l}{Storage (MiB)} \\ 
\toprule
AAC             & \multicolumn{3}{c}{312.07}        \\
Opus            & \multicolumn{3}{c}{163.23}        \\
NAF (Shared)  & \multicolumn{3}{c}{8.41}           \\
\bottomrule
\end{tabular}}
\caption{\small \textbf{Average space consumption across 6 Soundspaces scenes. Lower is better.}} 
\label{fig:storagetable}
\end{wraptable} 

%% file: figtext/learning_NAFs_test.tex
\begin{table*}[!t]
\footnotesize
\centering
\vspace{-2em}
\setlength{\tabcolsep}{1.5pt}
 \resizebox{1.0\linewidth}{!}{
 
\begin{tabular}{lllllllllllllllll}
	\toprule
	& \multicolumn{2}{c}{Large 1} & \multicolumn{2}{c}{Large 2} & \multicolumn{2}{c}{Medium 1} & \multicolumn{2}{c}{Medium 2} & \multicolumn{2}{c}{Small 1} & \multicolumn{2}{c}{Small 2} & \multicolumn{2}{c}{MeshRIR} & \multicolumn{2}{c}{Mean} \\ 
	\cmidrule(lr){2-3}  \cmidrule(lr){4-5} \cmidrule(lr){6-7} \cmidrule(lr){8-9} \cmidrule(lr){10-11} \cmidrule(lr){12-13} \cmidrule(lr){14-15} \cmidrule(lr){16-17}
	\bf Model                        & \addstackgap{Spec.$\downarrow$} & T60$\downarrow$ & Spec.$\downarrow$ & T60$\downarrow$ & Spec.$\downarrow$ & T60$\downarrow$ & Spec.$\downarrow$ & T60$\downarrow$ & Spec.$\downarrow$ & T60$\downarrow$ & Spec.$\downarrow$ & T60$\downarrow$ & Spec.$\downarrow$ & T60$\downarrow$ & Spec.$\downarrow$ & T60$\downarrow$ \\ 
	\midrule
	\addstackgap[1.5pt]{AAC-nearest} & 1.913                           & 9.996           & 1.989             & 13.31           & 2.111             & 6.148           & 2.122             & 6.051           & 2.296             & 9.798           & 2.509             & 5.809           & 1.057                 & 4.740               & 1.999             & 7.979           \\
		
	AAC-linear                       & 1.904                           & 8.847           & 1.964             & 11.63           & 2.105             & 4.585           & 2.116             & 4.422           & 2.299             & 8.253           & 2.521             & 6.021           & 1.081                 & 6.697               & 1.998             & 7.208\\
	Opus-nearest                     & 1.740                           & 12.20           & 1.817             & 15.15           & 1.887             & 7.875           & 1.898             & 7.897           & 2.058             & 10.68           & 2.238             & 7.564           & 1.711                 & 5.068               & 1.907             & 9.493           \\
	Opus-linear                      & 1.780                           & 11.30           & 1.827             & 13.55           & 1.922             & 6.710           & 1.934             & 6.917           & 2.097             & 9.116           & 2.284             & 6.981           & 1.743                 & 5.768               & 1.941             & 8.621           \\
	DSP                              & 1.106                           & 14.62           & 1.170             & 13.68           & 1.064             & 10.24           & 1.067             & 9.732           & 1.079             & 12.77           & 1.097             & 11.03           & N/A                 & N/A               & 1.097                 & 12.01           \\
	\midrule
	NAF (Dual)                       & 0.413                           & 6.288           & 0.421             & 7.111           & 0.386             & 3.173           & 0.387             & 3.169           & 0.365             & 3.497           & 0.361             & 2.210           & \textbf{0.403}                 & 4.201               & 0.388             & 4.241           \\
			
	NAF (Shared)                     & \textbf{0.396}                  & \textbf{4.166}  & \textbf{0.413}    & \textbf{6.075}  & \textbf{0.384}    & \textbf{3.110}  & \textbf{0.384}    & \textbf{3.072}  & \textbf{0.356}    & \textbf{3.378}  & \textbf{0.344}    & \textbf{2.098}  & \textbf{0.403}                 & \textbf{4.191}               & \textbf{0.380}    & \textbf{3.650}  \\
	\hline
\end{tabular}

}

\caption{\small \textbf{Quantitative Results on Test Set Accuracy.} We report the spectral loss between generated and ground truth log spectrograms across methods, as well as the percentage (\%) difference for the T60 reverberation time. The best method for each room is \textbf{bolded}. For the nearest and linear baselines, we perform interpolation in the time domain using samples from the training set. The DSP is not implemented for MeshRIR due to the lack of absolute room coordinates.}
\label{table:results}
\end{table*}

%% file: figtext/sparse_nerf_table.tex
\begin{table*}[!t]
\footnotesize
\setlength{\tabcolsep}{3.5pt}
\centering
\scalebox{0.9}{
\begin{tabular}{lcccccccccccc}
	\toprule
	& \multicolumn{6}{c}{Large Room 1} & \multicolumn{6}{c}{Large Room 2} \\ 
	\cmidrule(lr){2-7}  \cmidrule(lr){8-13} 
	& \multicolumn{3}{c}{PSNR $\uparrow$} & \multicolumn{3}{c}{SSIM $\uparrow$} & \multicolumn{3}{c}{PSNR $\uparrow$} & \multicolumn{3}{c}{SSIM $\uparrow$} \\ 
	\cmidrule(lr){2-4}  \cmidrule(lr){5-7} \cmidrule(lr){8-10} \cmidrule(lr){11-13}
	Training Images & 75             & 100            & 150            & 75             & 100            & 150            & 75             & 100            & 150            & 75             & 100            & 150            \\ 
	\midrule
	NeRF            & 25.41          & 27.36          & 29.85          & 0.872          & 0.892          & 0.926          & 25.70          & 27.74          & 29.34          & 0.821          & 0.853          & 0.879          \\
	NeRF + NAF      & \textbf{26.19} & \textbf{27.59} & \textbf{29.90} & \textbf{0.895} & \textbf{0.911} & \textbf{0.927} & \textbf{26.24} & \textbf{28.22} & \textbf{29.45} & \textbf{0.837} & \textbf{0.866} & \textbf{0.879} \\
	\bottomrule         
\end{tabular}
}
\vspace{-5pt}
\caption{\small \textbf{Quantitative Results on Cross-Modal Image Learning.} Quantitative results on joint training of NeRF and NAF jointly conditioned on a single local grid. We use very sparse training images in highly complex scenes. When evaluated on 50 test images, we observe that cross-modal learning helps improve PSNR and SSIM when the visual training data is more sparse. } 
 \label{fig:nerf}
 \vspace{-5pt}
\end{table*}

%% file: figtext/sparse_nerf.tex
\begin{figure}[H]
  \centering
  \includegraphics[width=1.0\linewidth]{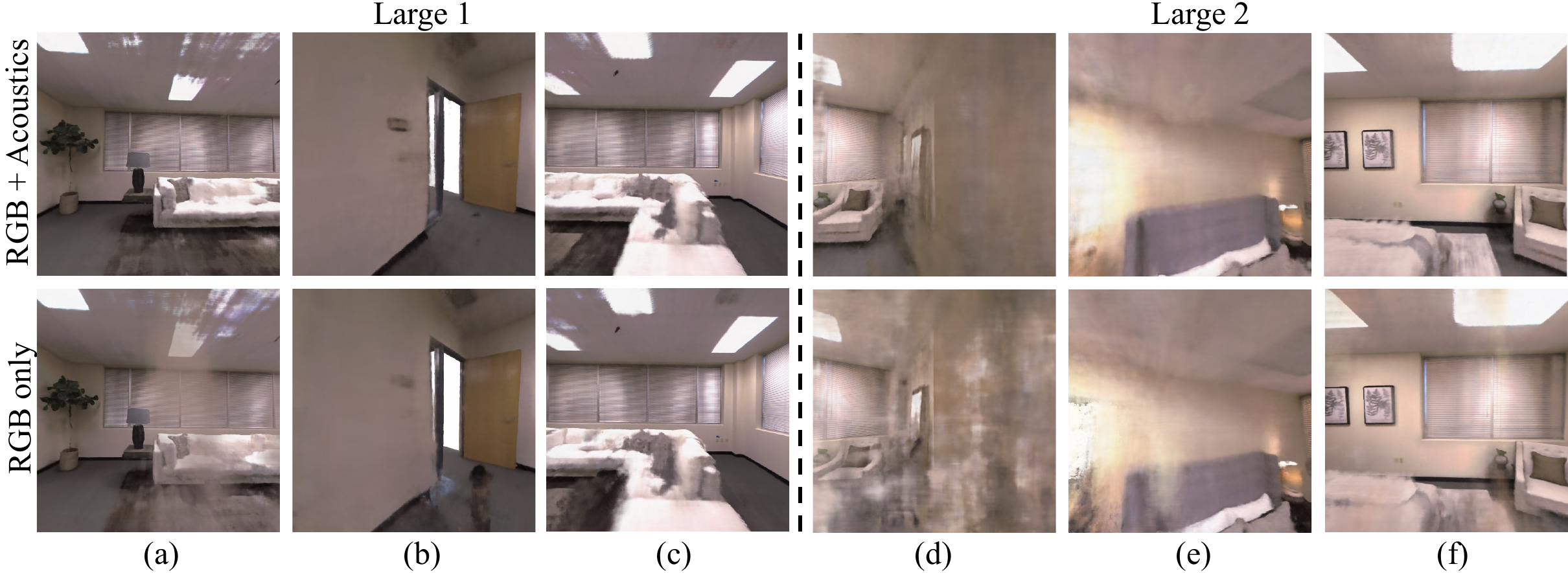}
  \vspace{-15pt}
  \caption{\small \textbf{Qualitative Visualization of Cross-Modal Image Learning.} Qualitative comparison between NeRF+NAFs with RGB and acoustic supervision, and NeRF learned with only RGB supervision. \textbf{(a)-(c)} Three views from "Large 1". \textbf{(d)-(f)} Three views from "Large 2". }
  \label{fig:sparse}
\end{figure}

%% file: figtext/ablation.tex
\begin{figure}[H]
\centering
  \includegraphics[width=0.8\linewidth]{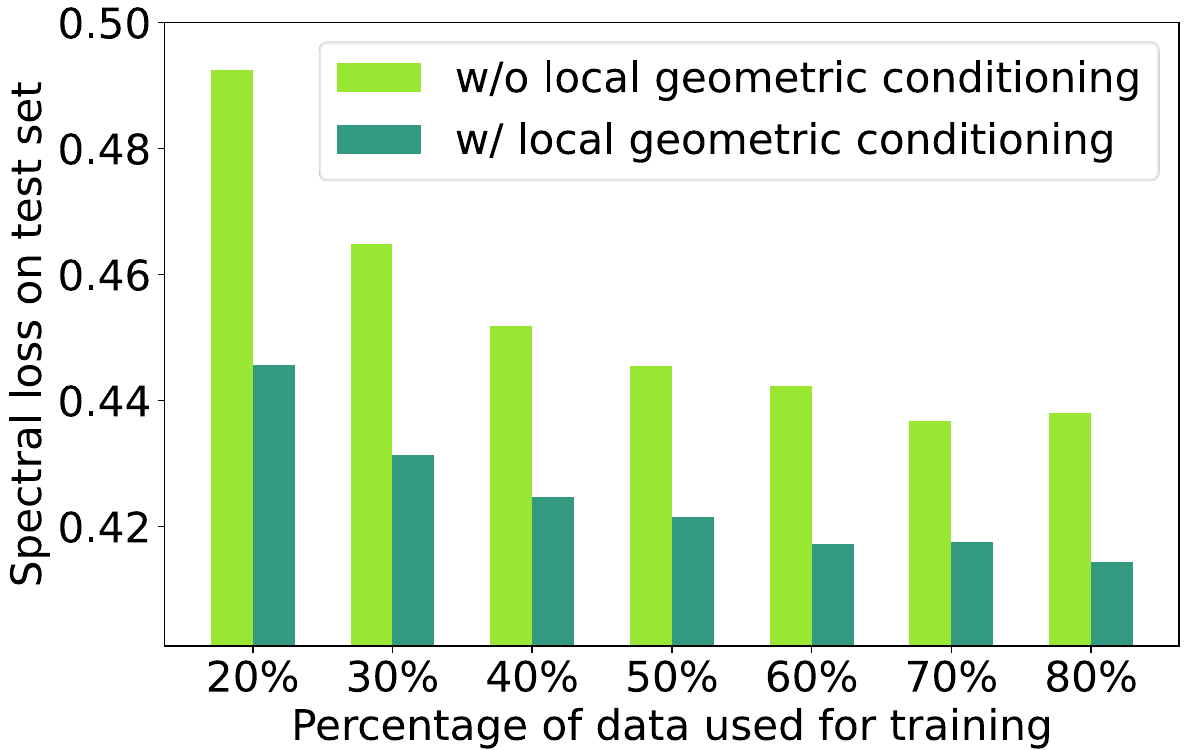}
  \caption{\small \textbf{Local Geometric Conditioning.} Comparison of NAF with and without local geometric conditioning trained with different amounts of data.}
  \label{fig:ablation}
\end{figure}

%% file: figtext/linear_probe_merged.tex
\begin{figure}[!t]
  \centering
  \includegraphics[width=0.85\linewidth]{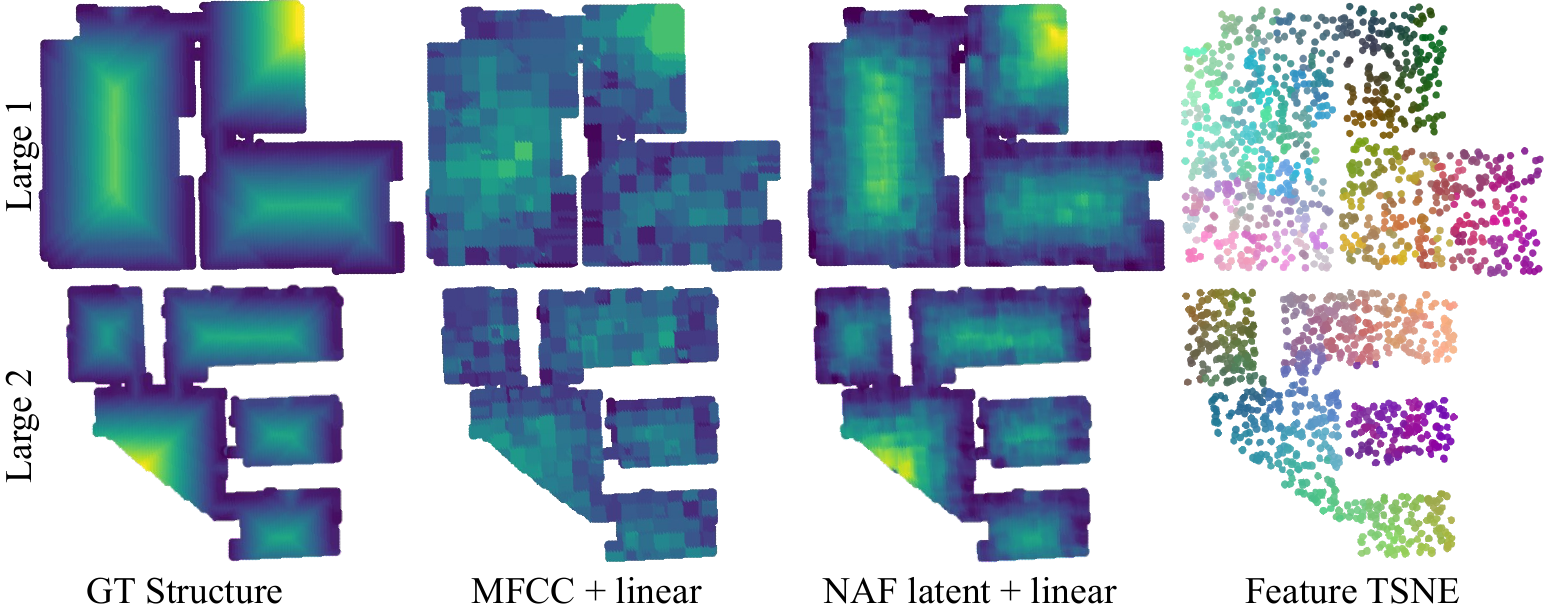}
  \caption{\textbf{Visualization of scene structure decoding with a linear layer.} \textbf{Column 1:} The ground truth scene structure map, at each position we visualize the distance to the nearest wall. \textbf{2:} Linear decoding results using MFCC features. \textbf{3:} Linear decoding results using NAF features. \textbf{4:} TSNE applied to the NAF features.}

  \label{fig:linearmap}
\end{figure} 

%% file: figtext/lin_probe_table.tex
\begin{table*}[!h]
\footnotesize
\setlength{\tabcolsep}{4.5pt}
\centering

\scalebox{1.0}{
\begin{tabular}{lclclclclclclcl}
\toprule
                  & \multicolumn{14}{c}{Explained variance}                                                                                                                                                                                                                          \\
                  
                  \cmidrule(lr){2-15} 
\textbf{Features} & \multicolumn{2}{c}{Large 1}        & \multicolumn{2}{c}{Large 2}        & \multicolumn{2}{c}{Medium 1}       & \multicolumn{2}{c}{Medium 2}       & \multicolumn{2}{c}{Small 1}        & \multicolumn{2}{c}{Small 2}        & \multicolumn{2}{c}{Mean}           \\ 
\midrule
MFCC              & \multicolumn{2}{c}{0.501}          & \multicolumn{2}{c}{0.458}          & \multicolumn{2}{c}{0.614}          & \multicolumn{2}{c}{0.642}          & \multicolumn{2}{c}{0.820}          & \multicolumn{2}{c}{0.723}          & \multicolumn{2}{c}{0.626}          \\
NAF latents       & \multicolumn{2}{c}{\textbf{0.908}} & \multicolumn{2}{c}{\textbf{0.891}} & \multicolumn{2}{c}{\textbf{0.900}} & \multicolumn{2}{c}{\textbf{0.923}} & \multicolumn{2}{c}{\textbf{0.936}} & \multicolumn{2}{c}{\textbf{0.916}} & \multicolumn{2}{c}{\textbf{0.913}} \\
\bottomrule
\end{tabular}
}
\caption{ \textbf{Quantitative Results on scene structure decoding.} We measure the explained variance scores of the predicted wall distance against the ground truth wall distance at test time locations after linear decoding. NAF latents consistently achieve higher explained variance scores than MFCC features.} 
 \label{fig:linprobetable}
\end{table*}

%% file: text/05_future.tex



%% file: text/06_conclusion.tex
\section{Discussion}
\label{sec:conclusion}
\myparagraph{Limitations and Future Work.} Although our method achieves generalization and high quality representations of acoustic fields within a single scene, NAFs do not currently generalize to multiple scenes.  Future work may explore generalization to novel scenes. One possible approaches may be to incorporate multi-modal inputs with the goal of synthesizing an acoustic field with few-shot visual or acoustic input.
\myparagraph{Societal impact.}
Our work focuses on learning a high quality representation of acoustic fields. The primary use case for our work lies in virtual reality and gaming. As our work can lead to more believable and higher quality representations of spatial audio than alternative methods, it is possible that our work could increase the dependency and time spent on gaming. The more compact nature of our acoustic representations may allow for spatial audio to be deployed to more systems, and enable more equitable access.

\myparagraph{Conclusion.} In summary, this paper introduces Neural Acoustic Fields (NAFs), a compact, continuous, and differentiable acoustic representation which can represent the underlying reverberation of different audio sources in a scene. By conditioning \model locally on the underlying scene geometry, we demonstrate that our approach enables the  prediction of plausible environmental reverberations even at unseen locations in the scene. Furthermore, we demonstrate that the acoustic representations learned through \model are powerful, and may be utilized to facilitate audio-visual cross-modal learning, as well as to infer the structure of scenes. 

%% file: text/09_acknowledgements.tex
\begin{ack}
 We would like to thank Leila Wehbe for providing feedback on our paper. This work was supported by the MIT-IBM Watson AI Lab, Amazon Research Award, ONR
MURI, DARPA Machine Common Sense program, ONR (N00014-18-1-2847), NIDCD R01 DC015988, Mitsubishi Electric, and the Tianqiao and Chrissy Chen Institute.
\end{ack}

%% file: appendix.tex
\newpage
\newcommand{\appendixhead}%
{\begin{center}\textbf{\Large Supplementary}\end{center}}

\appendixhead

\setcounter{table}{0}
\renewcommand{\thetable}{A\arabic{table}}
\setcounter{figure}{0}
\renewcommand{\thefigure}{A\arabic{figure}}
\appendix

\section{Additional Visualization of Rooms}

\input{figtext/supp_main}
We show additional NAF predictions of loudness as we move an emitter inside different rooms in Figure~\ref{fig:supp}. For each room, note how the sound is affected by the geometry. In wide open spaces the sound is highly dispersed. While in thin structures the sound tends to concentrate locally. As we move farther from the source, the loudness of the sound decreases.
\newpage
\section{Additional Visualization on Real-World Data}
\input{figtext/meshrir_text}
In Figure~\ref{fig:meshrir} we compare on the MeshRIR dataset which is collected from the real-world. Bilinear interpolation introduces characteristic artifacts at the sample boundaries, while nearest neighbor has discretization artifacts. In contrast, our NAFs are able to predict a smoothly varying acoustic field despite learning from discretely sampled training data.
\newpage
\section{Additional quantitative results}
\input{figtext/Helmholtz}
In Table~\ref{fig:Helmholtz}, we compare our method against "Kernel Ridge Regression with Constraint of Helmholtz Equation for Sound Field Interpolation" on the MeshRIR dataset. "Ridge-Orig" denotes the authors proposed setup which applies a 500Hz low pass filter. While "Ridge-Unfiltered" is a modified setup where we do not perform a low pass. Note that their method requires an individual model for each unique emitter location, while our NAFs can be queried using any emitter/receiver position.

\input{figtext/IACC}
In Table~\ref{fig:IACC_scores} we evaluate the error in the interaural cross correlation coefficient (IACC). The IACC is correlated with the ability for humans to localize a sound. We find that NAFs have low IACC error.

\newpage
\section{Architecture and Training Details}

We visualize all three models that we experiment with.

In Figure~\ref{fig:dual} is a network that uses different local feature grids for the emitter and receiver (dual grids). The network uses the emitter and listener positions to sample from the two different grids. 

In Figure~\ref{fig:netarch_supp} we show a model where the local feature grids for the emitter and receiver are shared. This network uses the emitter and listener positions to sample from the same shared grid.

In Figure~\ref{fig:no} we show a model that does not utilize any kind of local geometry conditioning. 

The listener, emitter, phase, and time input are transformed using sinusoidal embedding, while the orientation and left/right are retrieved. All transformed inputs are directly fed to the network. We find that the sharing the feature grid  performs better than using different local feature grids.

\input{figtext/dual_grid}
\input{figtext/arch_supp}

\input{figtext/nogrid}

Each network consists of 8 fully connected layers in a feedforward fashion, as well as a skip connection consisting of two fully connected layers. The skip connection takes the input and adds its output to that for the fourth intermediate layer. We utilize an intermediate feature size of 512, and Leaky ReLU with a slope of 0.1 as the activation function. The grid is initialized to stretch the bounding box of a scene. Each point is located at a distance of 0.25m from the nearest neighbor. 64 features are used for each point. Each element of the grid is initialized i.i.d. from $\mathcal{N}(0, \frac{1}{\sqrt{64}})$. We initialize the bandwidth for each point at $\sigma=0.25$, and jointly train the bandwidth as part of the network. 
For the network and the grid, we utilize an initial learning rate of $5e-4$. The \textit{Adam} optimizer is used when training our network. We utilize a orientation embedding of shape $\mathcal{R}^{7 \times 4 \times 512}$ where $7$ is the number of intermediate outputs, $4$ is the number of orientations, and $512$ is the feature dimension. For the left-right embedding, we use a shape of $\mathcal{R}^{7 \times 2 \times 512}$. We perform additive conditioning by adding a $\mathcal{R}^{512}$ vector to each intermediate output for both the orientation and the left/right.

For each scene, to generate a log-spectrogram for each impulse response, we compute the mean and standard deviation  $\mu_{(t,f)}, \sigma_{(t,f)}$ for each time/frequency index in the log-spectrogram, and normalize the data prior to training:
\begin{align*}
   \vv_\text{STFT\_mag}(t, f) = \frac{\vv_\text{STFT\_mag}(t, f) -\mu_{(t,f)}}{3.0 \times \sigma_{(t,f)}}
\end{align*}

To generate the instantaneous frequency (phase) representation for each impulse response, we normalize the data prior to training:
\begin{align*}
   \vv_\text{STFT\_IF} = \frac{\vv_\text{STFT\_IF}}{3.0 \times \sigma_\text{IF}}
\end{align*}

For the sinusoidal embedding, we utilize both $\cos$ and $\sin$ with $10$ frequencies each for encoding position, phase, and time. For encoding position we utilize a max frequency of $2^{7}$Hz, while for encoding time and frequency we utilize a max frequency of $2^{10}$Hz. 

Since we do not know beforehand the time duration of an impulse response at an unseen location, we compute the maximum impulse length for each scene and use this length to zero pad the training impulse responses. Because the padded regions do not contain useful information, we want the network to focus modeling efforts on the early regions of the impulse response. We achieve this by stochastically padding the impulse response to maximum impulse length with $0.1$ probability. Because the implicit function is trained on individual $(t,f)$ coordinates within a given $v_{\text{STFT}}$, training samples do not need to be of the same length.
During test time, we perform inference up to the maximum duration of scene impulse response.

\section{Dataset Visualization} 
\input{figtext/probes} 
In Figure~\ref{fig:myprobes}, we visualize both the room and  underlying set of probe positions in the training data.  Due to occlusion and the geometry, even slightly moving the emitter or listener position can result in different results. As we demonstrated, both nearest neighbor and linear interpolation perform poorly compared to our learned solution. In contrast, recovered acoustic fields from NAF trained on these probe positions is substantially denser (Figure~\ref{fig:supp}).

\section {Storage Comparison}
\input{figtext/average_storage}
We compare the averaged on disk storage cost of the different methods for inferring the spatial audio using a precomputed training set in Table~\ref{fig:storage}. Both linear and nearest interpolation methods require access to the entire training set, while our NAF based approaches compactly encode the acoustic scene.

\section{Details of the compression baselines}
\label{sec:compression}
If uncompressed, the precomputed spatial acoustic field can reach gigabyte or terabyte sizes depending on probe density, scene size, and bandwidth of the impulse.
When applied to gaming and virtual reality applications, minimizing the space taken up by these acoustic representations is critical and have been widely studied. 

We utilize two state-of-the-art lossy coding methods applied to the audio. They are respectively Advanced Audio Coding (AAC-LC) and Xiph Opus. These two methods were chosen because they are in widespread usage for media encoding, are among the best coding methods for a given bitrate, and have high quality open-source implementations available. The bitrates were selected on the basis of attempting to match the size of the NAFs representations, while being allowed by the respective encoders.

We describe the parameters and additional details for these two coding methods.

\subsection{AAC baseline}
We utilize \texttt{ffmpeg 5.0}, and select the open source "aac" implementation. We set the combined stereo bitrate to 24 kBit/s (12kBit/s per channel) in constant bit rate mode, as we found that there are occasional encode/decode failures below this bitrate. 

\subsection{Opus baseline}
We utilize \texttt{opustools 0.2} backed by \texttt{libopus 1.3.1}. The encoder is set to 12kBit/s for stereo (6kBit/s per channel) in constrained variable bitrate mode. Complexity it set to the maximum of 10, and music mode is set (as opposed to speech tuning mode).

\section{{Alternative Neural Representations}}
\input{figtext/wave_phase}
Our current method follows prior work in learning the log-magnitude STFT and instantaneous frequency phase. In this section, we investigate a possible alternative of  directly learning in the time domain. The MSE and T60 error percentage is presented in Table~\ref{fig:wavephase}. We observe that  modeling in the time domain performs poorly.

\section{{$L_2$ regularized grid in NeRF}}

\input{figtext/l2_reg}
In Table~\ref{fig:l2reg} we compare NeRF that utilizes a grid and trained using image reconstruction loss, against a variant where a $L_2$ penalty with weight $1e-5$ to ensure a smooth latent space is added to the image reconstruction loss. There are 75 images used in the training set. We observe degraded performance when we apply this penalty. This indicates that our NAFs are providing more information than simple regularization to ensure a smooth latent grid.

%% file: figtext/supp_main.tex
\begin{figure}[!h]
  \centering
  \includegraphics[width=0.95\linewidth]{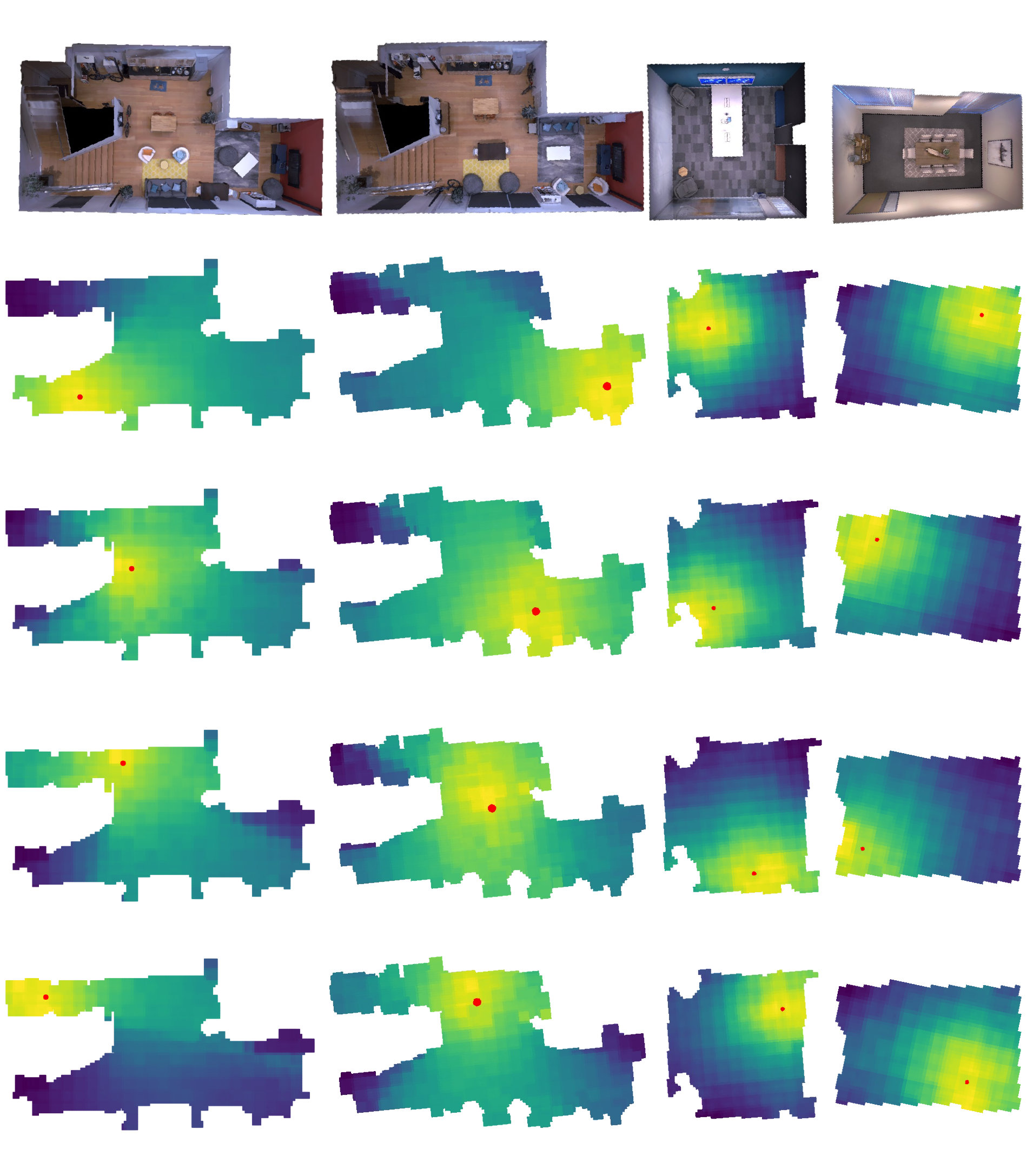}
  \vspace{-0.5em}
  \caption{\textbf{Additional Qualitative Predictions of NAF.} Qualitative visualization of the loudness map as predicted by NAF across four different rooms.}
 \vspace{-1mm}
  \label{fig:supp}
\end{figure}

%% file: figtext/meshrir_text.tex
\begin{figure}[!h]
  \centering
  \includegraphics[width=1.0\linewidth]{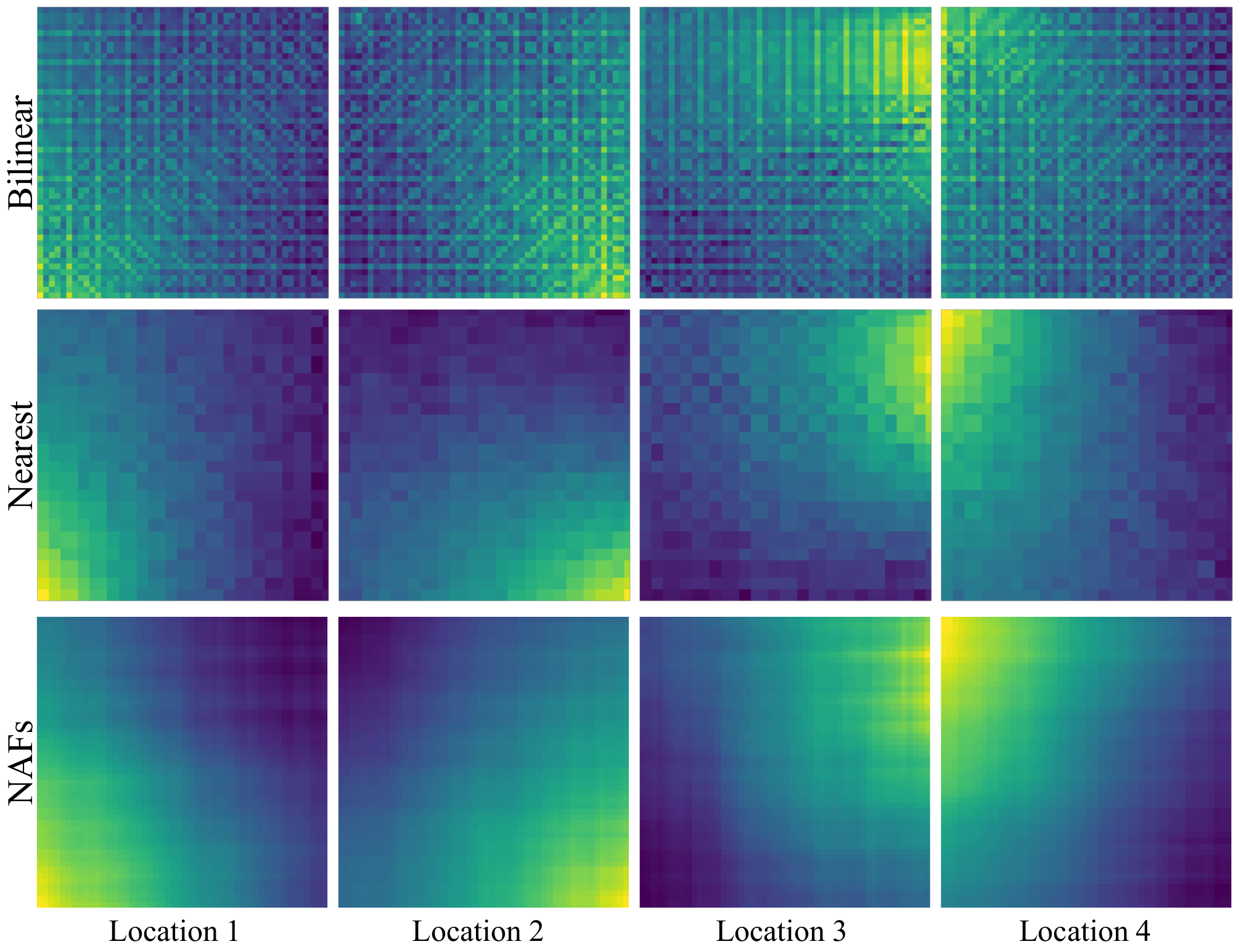}
   \vspace{-3mm}
  \caption{\textbf{Comparison on the MeshRIR real-world dataset.} We compare our method on the MeshRIR dataset across four emitter locations. \textbf{Top.} The loudness map using bilinear interpolation of the ground truth. \textbf{Middle.} The loudness map using nearest interpolation of the ground truth. \textbf{Bottom.} The loudness map predicted using NAFs. Our method can predict a smoothly varying loudness map without artifacts.}
  \label{fig:meshrir}
\end{figure}

%% file: figtext/Helmholtz.tex
\begin{table*}[!h]
\footnotesize
\setlength{\tabcolsep}{3.7pt}
\centering

{\begin{tabular}{llll}
\hline
                     & Spectral  & T60       & DRR       \\ \hline
Ridge-Orig     & 2.539     & 8.192     & 2.497     \\
Ridge-Unfiltered & 1.370     & 6.294     & 3.702     \\ \hline
NAF (Dual)           & \textbf{0.403}     & 4.201     & 0.992     \\
NAF (Shared)         & \textbf{0.403} & \textbf{4.191} & \textbf{0.972} \\ \hline
\end{tabular}}
\caption{{\textbf{Comparison against a kernel regression baseline} We compare against a kernel ridge regression baseline on the MeshRIR dataset. We find that our NAFs perform better on the metrics evaluated.}}
 \label{fig:Helmholtz}
\end{table*}

%% file: figtext/IACC.tex
\begin{table*}[!h]
\footnotesize
\setlength{\tabcolsep}{3.7pt}
\centering

{\begin{tabular}{llllllll}
\hline
\multicolumn{8}{c}{IACC error $\downarrow$}                                                                \\
Method       & Large 1   & Large 2   & Medium 1  & Medium 2  & Small 1   & Small 2   & Mean      \\ \hline
AAC-nearest  & 236.8     & 184.2     & 213.7     & 215.3     & 264.8     & 272.5     & 231.2     \\
AAC-linear   & 212.3     & 156.7     & 185.9     & 187.8     & 245.2     & 265.2     & 208.8     \\
Opus-nearest & 73.75     & 45.97     & 71.97     & 74.70     & 103.8     & \textbf{67.40} & 72.93     \\
Opus-linear  & 75.56     & 48.32     & 73.38     & 77.33     & 109.2     & 78.10     & 76.98     \\
DSP          & 460.5     & 446.0     & 430.0     & 430.1     & 443.6     & 446.3     & 442.7     \\ \hline
NAF (Dual) & {74.01} & {45.94} & {71.89} & {74.70} & {103.8} & \textbf{67.40} & {72.96}\\
NAF (Shared) & \textbf{73.68} & \textbf{45.90} & \textbf{71.52} & \textbf{73.58} & \textbf{103.6} & \textbf{67.40} & \textbf{72.62}
\end{tabular}}
\caption{{\textbf{Mean absolute error of IACC.} We compute interaural cross correlation coefficient (IACC) using the impulse response from the left and right ears. Here we show the mean absolute error of the IACC for a given method and the ground truth. Units are seconds, for visualization values are multiplied by $1\mathrm{e}6$, lower is better.}}
 \label{fig:IACC_scores}
\end{table*}

%% file: figtext/dual_grid.tex
\begin{figure}[!h]
  \centering
  \includegraphics[width=0.9\linewidth]{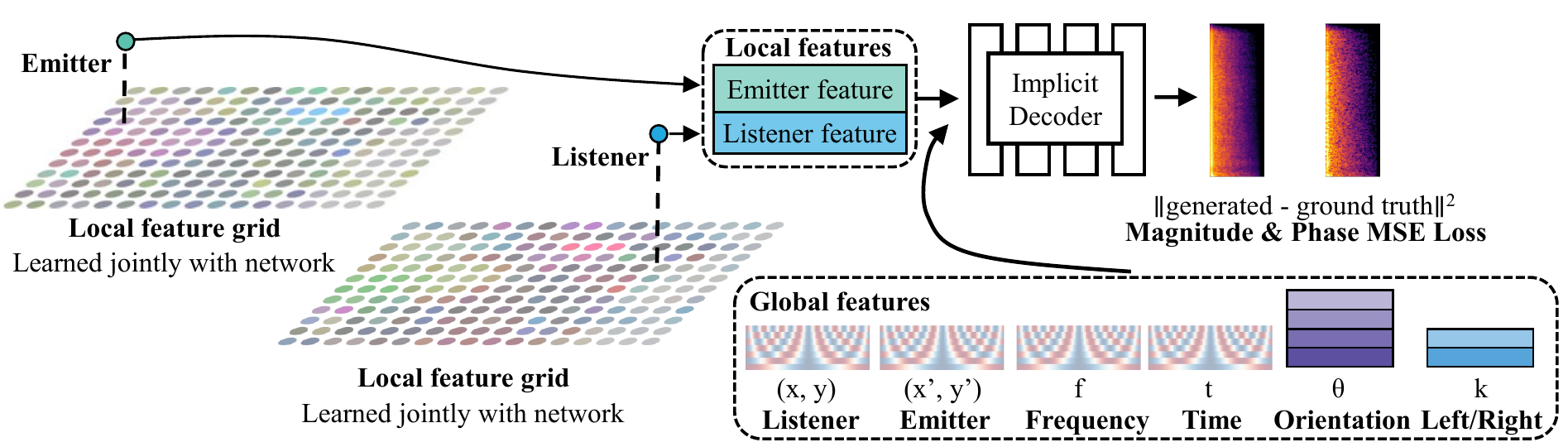}
  \vspace{-0.5em}
  \caption{Architecture of the model that uses emitter and listener specific local geometry conditioning.}
 \vspace{-3mm}
  \label{fig:dual}
\end{figure}

%% file: figtext/arch_supp.tex
\begin{figure*}[!h]
  \centering
  \includegraphics[width=0.84\linewidth]{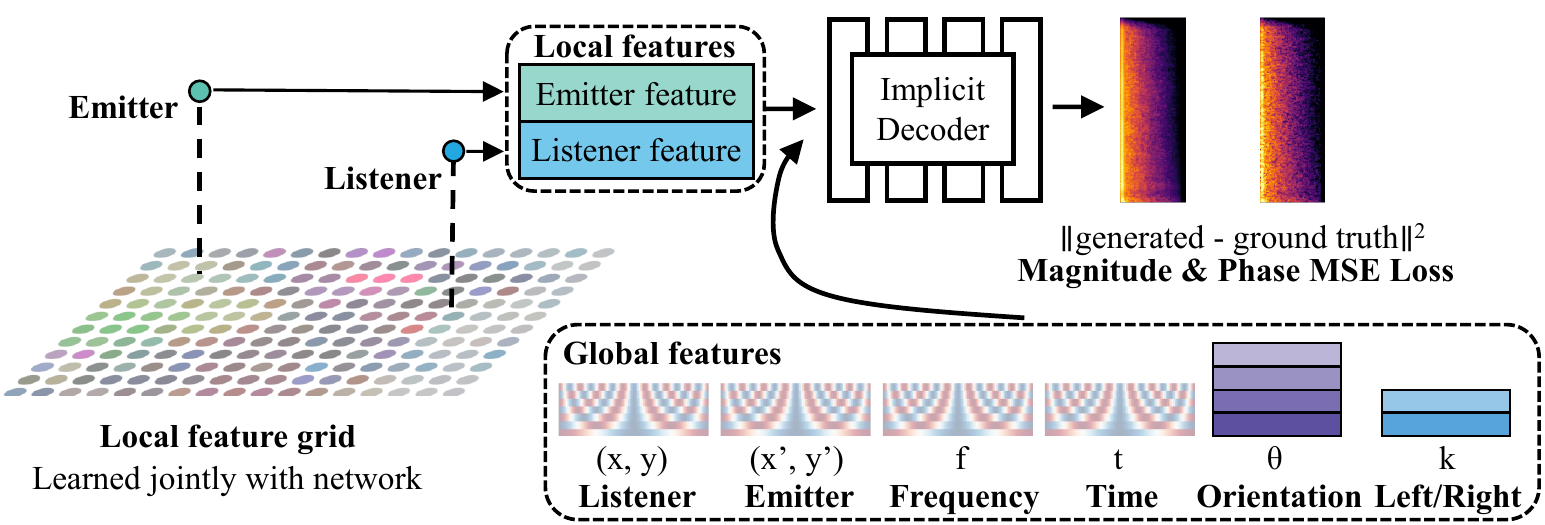}
\vspace{-5pt}
  \caption{\small Architecture of the model that share emitter and listener local geometry conditioning.}
 \vspace{-10pt}
  \label{fig:netarch_supp}
\end{figure*} 

%% file: figtext/nogrid.tex
\begin{figure}[!h]
  \centering
  \includegraphics[width=0.6\linewidth]{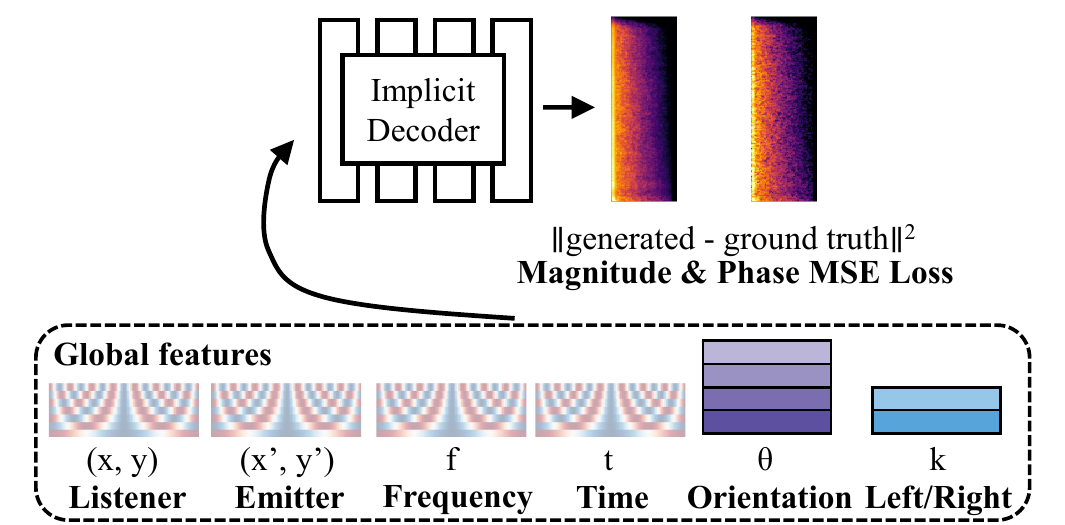}
  \vspace{-0.5em}
  \caption{Architecture of the model that uses no local geometry conditioning.}
 \vspace{-3mm}
  \label{fig:no}
\end{figure}

%% file: figtext/probes.tex
\begin{figure}[!ht]
  \centering
  \includegraphics[width=1.0\linewidth]{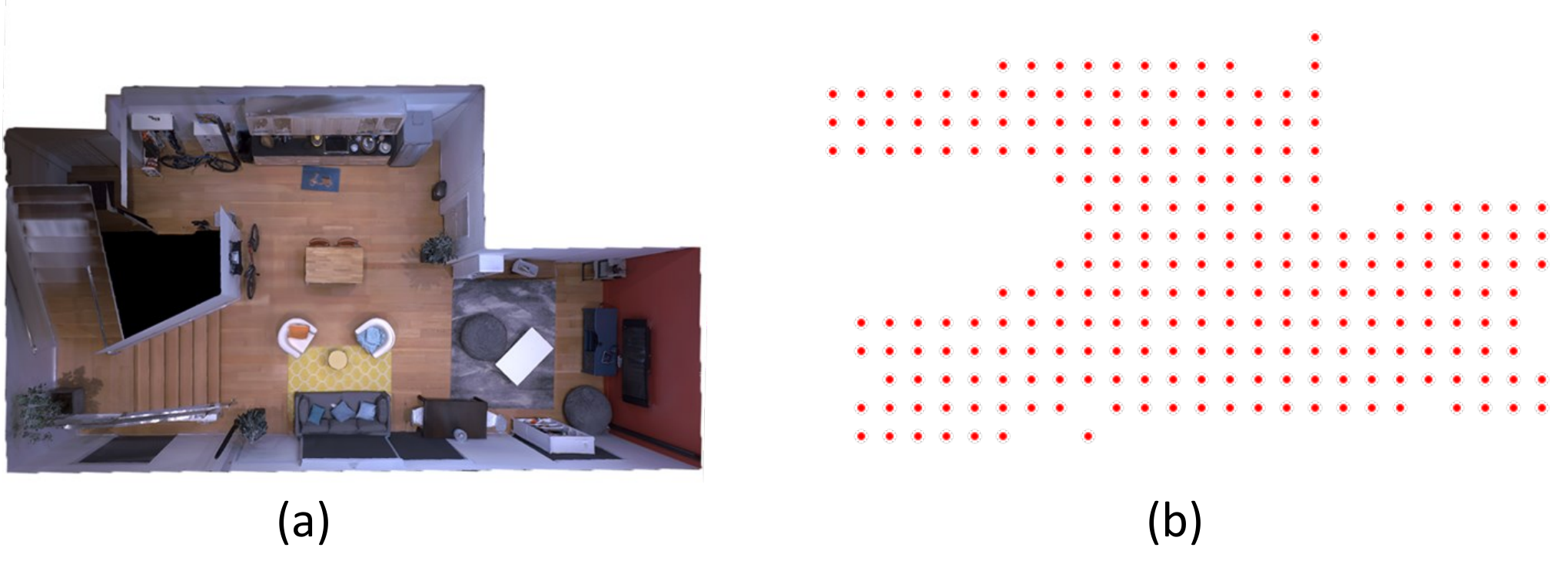}
  \vspace{-0.5em}
  \caption{A room the emitter-listener probes. \textbf{(a)} The 3D structure of a room. \textbf{b} The probes marking the location of emitters/listeners.}
 \vspace{-3mm}
  \label{fig:myprobes}
\end{figure}

%% file: figtext/average_storage.tex
\begin{table*}[!h]
\footnotesize
\setlength{\tabcolsep}{3.5pt}
\centering

\begin{tabular}{lccccccc}
\toprule
                & \multicolumn{7}{c}{Storage (MiB)}                                    \\
\textbf{Method} & Large 1 & Large 2 & Medium 1 & Medium 2 & Small 1 & Small 2 & Mean   \\ \hline
AAC             & 495.97  & 478.55  & 483.42   & 451.14   & 116.75  & 54.64   & 346.74 \\
Opus            & 258.51  & 257.08  & 245.65   & 231.06   & 66.15   & 29.75   & 181.37 \\
NAF (Dual)      & 8.78    & 8.87    & 8.87     & 8.92     & 8.45    & 8.37    & 8.71   \\
NAF (Shared)    & 8.44    & 8.49    & 8.49     & 8.51     & 8.28    & 8.23    & 8.41   \\ \bottomrule
\end{tabular}
\caption{{\textbf{Total storage cost of different methods.} We average the amount of data required for different methods for the six scenes. Our NAFs are able to compactly represent the scene while maintaining higher quality.}} 
 \label{fig:storage}
\end{table*}

%% file: figtext/wave_phase.tex
\begin{table*}[!h]
\footnotesize
\setlength{\tabcolsep}{3.5pt}
\centering

\begin{tabular}{lcc}
\hline
{Representation}    & {Spectral loss $\downarrow$}   & {T60$\downarrow$}   \\ \hline
{Time domain}       & {2.046} & {49.72}\\
{NAFs} & {0.396} & {4.166} \\ \hline
\end{tabular}
\caption{{\textbf{Learning different representations} We compare NAFs in the STFT domain against directly learning in the time domain.}} 
 \label{fig:wavephase}
\end{table*}

%% file: figtext/l2_reg.tex
\begin{table*}[!h]
\footnotesize
\setlength{\tabcolsep}{3.5pt}
\centering

\begin{tabular}{lcccc}
\hline
         & \multicolumn{2}{c}{{Large 1}} & \multicolumn{2}{c}{{Large 2}} \\
              & {PSNR} $\uparrow$         & {MSE} $\downarrow$         & {PSNR} $\uparrow$        & {MSE}  $\downarrow$        \\ \hline
{NeRF + grid + $L_{2}$} & {22.69}         & {6.956}        & {24.86}        & {7.128}        \\
{NeRF + grid}          & {\textbf{25.41}}         & {\textbf{6.618}}        & {\textbf{25.70}}        & {\textbf{6.921}}       
\end{tabular}
\caption{{\textbf{Regularizing the grid.} In this experiment, we compare learning NeRF with a grid without regularization, and with $L_2$ regularization.}} 
 \label{fig:l2reg}
\end{table*}

%% file: main_NIPS.bbl
\begin{thebibliography}{39}
\providecommand{\natexlab}[1]{#1}
\providecommand{\url}[1]{\texttt{#1}}
\expandafter\ifx\csname urlstyle\endcsname\relax
  \providecommand{\doi}[1]{doi: #1}\else
  \providecommand{\doi}{doi: \begingroup \urlstyle{rm}\Url}\fi

\bibitem[Sitzmann et~al.(2019)Sitzmann, Zollh{\"o}fer, and
  Wetzstein]{sitzmann2019srns}
Vincent Sitzmann, Michael Zollh{\"o}fer, and Gordon Wetzstein.
\newblock Scene representation networks: Continuous 3d-structure-aware neural
  scene representations.
\newblock In \emph{Proc. NeurIPS 2019}, 2019.

\bibitem[Mildenhall et~al.(2020)Mildenhall, Srinivasan, Tancik, Barron,
  Ramamoorthi, and Ng]{mildenhall2020nerf}
Ben Mildenhall, Pratul~P Srinivasan, Matthew Tancik, Jonathan~T Barron, Ravi
  Ramamoorthi, and Ren Ng.
\newblock Nerf: Representing scenes as neural radiance fields for view
  synthesis.
\newblock In \emph{Proc. ECCV}, 2020.

\bibitem[Niemeyer et~al.(2020)Niemeyer, Mescheder, Oechsle, and
  Geiger]{Niemeyer2020DVR}
Michael Niemeyer, Lars Mescheder, Michael Oechsle, and Andreas Geiger.
\newblock Differentiable volumetric rendering: Learning implicit 3d
  representations without 3d supervision.
\newblock In \emph{Proc. CVPR}, 2020.

\bibitem[Yariv et~al.(2020)Yariv, Kasten, Moran, Galun, Atzmon, Ronen, and
  Lipman]{yariv2020multiview}
Lior Yariv, Yoni Kasten, Dror Moran, Meirav Galun, Matan Atzmon, Basri Ronen,
  and Yaron Lipman.
\newblock Multiview neural surface reconstruction by disentangling geometry and
  appearance.
\newblock \emph{Proc. NeurIPS}, 2020.

\bibitem[Raghuvanshi and Snyder(2014)]{raghuvanshi2014parametric}
Nikunj Raghuvanshi and John Snyder.
\newblock Parametric wave field coding for precomputed sound propagation.
\newblock \emph{ACM Transactions on Graphics (TOG)}, 33\penalty0 (4):\penalty0
  1--11, 2014.

\bibitem[Raghuvanshi and Snyder(2018)]{raghuvanshi2018parametric}
Nikunj Raghuvanshi and John Snyder.
\newblock Parametric directional coding for precomputed sound propagation.
\newblock \emph{ACM Transactions on Graphics (TOG)}, 37\penalty0 (4):\penalty0
  1--14, 2018.

\bibitem[Chaitanya et~al.(2020)Chaitanya, Raghuvanshi, Godin, Zhang,
  Nowrouzezahrai, and Snyder]{chaitanya2020directional}
Chakravarty R~Alla Chaitanya, Nikunj Raghuvanshi, Keith~W Godin, Zechen Zhang,
  Derek Nowrouzezahrai, and John~M Snyder.
\newblock Directional sources and listeners in interactive sound propagation
  using reciprocal wave field coding.
\newblock \emph{ACM Transactions on Graphics (TOG)}, 39\penalty0 (4):\penalty0
  44--1, 2020.

\bibitem[Srinivasan et~al.(2021)Srinivasan, Deng, Zhang, Tancik, Mildenhall,
  and Barron]{nerv2021}
Pratul~P. Srinivasan, Boyang Deng, Xiuming Zhang, Matthew Tancik, Ben
  Mildenhall, and Jonathan~T. Barron.
\newblock Nerv: Neural reflectance and visibility fields for relighting and
  view synthesis.
\newblock In \emph{Proc. CVPR}, 2021.

\bibitem[Mignot et~al.(2013)Mignot, Chardon, and Daudet]{mignot2013low}
R{\'e}mi Mignot, Gilles Chardon, and Laurent Daudet.
\newblock Low frequency interpolation of room impulse responses using
  compressed sensing.
\newblock \emph{IEEE/ACM Transactions on Audio, Speech, and Language
  Processing}, 22\penalty0 (1):\penalty0 205--216, 2013.

\bibitem[Antonello et~al.(2017)Antonello, De~Sena, Moonen, Naylor, and
  Van~Waterschoot]{antonello2017room}
Niccolo Antonello, Enzo De~Sena, Marc Moonen, Patrick~A Naylor, and Toon
  Van~Waterschoot.
\newblock Room impulse response interpolation using a sparse spatio-temporal
  representation of the sound field.
\newblock \emph{IEEE/ACM Transactions on Audio, Speech, and Language
  Processing}, 25\penalty0 (10):\penalty0 1929--1941, 2017.

\bibitem[Ueno et~al.(2018)Ueno, Koyama, and Saruwatari]{ueno2018kernel}
Natsuki Ueno, Shoichi Koyama, and Hiroshi Saruwatari.
\newblock Kernel ridge regression with constraint of helmholtz equation for
  sound field interpolation.
\newblock In \emph{2018 16th International Workshop on Acoustic Signal
  Enhancement (IWAENC)}, pages 1--440. IEEE, 2018.

\bibitem[Mehra et~al.(2014)Mehra, Antani, Kim, and Manocha]{mehra2014source}
Ravish Mehra, Lakulish Antani, Sujeong Kim, and Dinesh Manocha.
\newblock Source and listener directivity for interactive wave-based sound
  propagation.
\newblock \emph{IEEE transactions on visualization and computer graphics},
  20\penalty0 (4):\penalty0 495--503, 2014.

\bibitem[Ratnarajah et~al.(2021)Ratnarajah, Zhang, Yu, Tang, Manocha, and
  Yu]{ratnarajah2021fast}
Anton Ratnarajah, Shi-Xiong Zhang, Meng Yu, Zhenyu Tang, Dinesh Manocha, and
  Dong Yu.
\newblock Fast-rir: Fast neural diffuse room impulse response generator.
\newblock \emph{arXiv preprint arXiv:2110.04057}, 2021.

\bibitem[Richard et~al.(2020)Richard, Markovic, Gebru, Krenn, Butler, Torre,
  and Sheikh]{richard2020neural}
Alexander Richard, Dejan Markovic, Israel~D Gebru, Steven Krenn,
  Gladstone~Alexander Butler, Fernando Torre, and Yaser Sheikh.
\newblock Neural synthesis of binaural speech from mono audio.
\newblock In \emph{International Conference on Learning Representations}, 2020.

\bibitem[Richard et~al.(2022)Richard, Dodds, and Ithapu]{richard2022deep}
Alexander Richard, Peter Dodds, and Vamsi~Krishna Ithapu.
\newblock Deep impulse responses: Estimating and parameterizing filters with
  deep networks.
\newblock In \emph{ICASSP 2022-2022 IEEE International Conference on Acoustics,
  Speech and Signal Processing (ICASSP)}, pages 3209--3213. IEEE, 2022.

\bibitem[Niemeyer et~al.(2019)Niemeyer, Mescheder, Oechsle, and
  Geiger]{Niemeyer2019ICCV}
Michael Niemeyer, Lars Mescheder, Michael Oechsle, and Andreas Geiger.
\newblock Occupancy flow: 4d reconstruction by learning particle dynamics.
\newblock In \emph{Proc. ICCV}, 2019.

\bibitem[Chen and Zhang(2019)]{chen2019learning}
Zhiqin Chen and Hao Zhang.
\newblock Learning implicit fields for generative shape modeling.
\newblock In \emph{Proc. CVPR}, pages 5939--5948, 2019.

\bibitem[Park et~al.(2019)Park, Florence, Straub, Newcombe, and
  Lovegrove]{park2019deepsdf}
Jeong~Joon Park, Peter Florence, Julian Straub, Richard Newcombe, and Steven
  Lovegrove.
\newblock Deepsdf: Learning continuous signed distance functions for shape
  representation.
\newblock In \emph{Proc. CVPR}, 2019.

\bibitem[Saito et~al.(2019)Saito, Huang, Natsume, Morishima, Kanazawa, and
  Li]{saito2019pifu}
Shunsuke Saito, Zeng Huang, Ryota Natsume, Shigeo Morishima, Angjoo Kanazawa,
  and Hao Li.
\newblock Pifu: Pixel-aligned implicit function for high-resolution clothed
  human digitization.
\newblock In \emph{Proc. ICCV}, pages 2304--2314, 2019.

\bibitem[Hong et~al.(2022)Hong, Du, Lin, Tenenbaum, and Gan]{hong20223d}
Yining Hong, Yilun Du, Chunru Lin, Joshua~B Tenenbaum, and Chuang Gan.
\newblock 3d concept grounding on neural fields.
\newblock \emph{arXiv preprint arXiv:2207.06403}, 2022.

\bibitem[Wang et~al.(2021)Wang, Wang, Genova, Srinivasan, Zhou, Barron,
  Martin-Brualla, Snavely, and Funkhouser]{wang2021ibrnet}
Qianqian Wang, Zhicheng Wang, Kyle Genova, Pratul~P Srinivasan, Howard Zhou,
  Jonathan~T Barron, Ricardo Martin-Brualla, Noah Snavely, and Thomas
  Funkhouser.
\newblock Ibrnet: Learning multi-view image-based rendering.
\newblock In \emph{Proceedings of the IEEE/CVF Conference on Computer Vision
  and Pattern Recognition}, pages 4690--4699, 2021.

\bibitem[Jiang et~al.(2020)Jiang, Sud, Makadia, Huang, Nie{\ss}ner, and
  Funkhouser]{jiang2020local}
Chiyu Jiang, Avneesh Sud, Ameesh Makadia, Jingwei Huang, Matthias Nie{\ss}ner,
  and Thomas Funkhouser.
\newblock Local implicit grid representations for 3d scenes.
\newblock In \emph{Proc. CVPR}, pages 6001--6010, 2020.

\bibitem[DeVries et~al.(2021)DeVries, Bautista, Srivastava, Taylor, and
  Susskind]{devries2021unconstrained}
Terrance DeVries, Miguel~Angel Bautista, Nitish Srivastava, Graham~W Taylor,
  and Joshua~M Susskind.
\newblock Unconstrained scene generation with locally conditioned radiance
  fields.
\newblock \emph{arXiv preprint arXiv:2104.00670}, 2021.

\bibitem[Aytar et~al.(2016)Aytar, Vondrick, and Torralba]{aytar2016soundnet}
Yusuf Aytar, Carl Vondrick, and Antonio Torralba.
\newblock Soundnet: Learning sound representations from unlabeled video.
\newblock \emph{Advances in neural information processing systems},
  29:\penalty0 892--900, 2016.

\bibitem[Arandjelovic and Zisserman(2017)]{arandjelovic2017look}
Relja Arandjelovic and Andrew Zisserman.
\newblock Look, listen and learn.
\newblock In \emph{Proceedings of the IEEE International Conference on Computer
  Vision}, pages 609--617, 2017.

\bibitem[Senocak et~al.(2018)Senocak, Oh, Kim, Yang, and
  Kweon]{senocak2018learning}
Arda Senocak, Tae-Hyun Oh, Junsik Kim, Ming-Hsuan Yang, and In~So Kweon.
\newblock Learning to localize sound source in visual scenes.
\newblock In \emph{Proceedings of the IEEE Conference on Computer Vision and
  Pattern Recognition}, pages 4358--4366, 2018.

\bibitem[Zhao et~al.(2018)Zhao, Gan, Rouditchenko, Vondrick, McDermott, and
  Torralba]{zhao2018sound}
Hang Zhao, Chuang Gan, Andrew Rouditchenko, Carl Vondrick, Josh McDermott, and
  Antonio Torralba.
\newblock The sound of pixels.
\newblock In \emph{Proceedings of the European conference on computer vision
  (ECCV)}, pages 570--586, 2018.

\bibitem[Chen et~al.(2020)Chen, Jain, Schissler, Gari, Al-Halah, Ithapu,
  Robinson, and Grauman]{chen2020soundspaces}
Changan Chen, Unnat Jain, Carl Schissler, Sebastia Gari, Ziad Al-Halah,
  Vamsi~Krishna Ithapu, Philip Robinson, and Kristen Grauman.
\newblock Soundspaces: Audio-visual navigation in 3d environments.
\newblock In \emph{ECCV}, 2020.

\bibitem[Singh et~al.(2021)Singh, Mentch, Ng, Beveridge, and
  Drori]{singh2021image2reverb}
Nikhil Singh, Jeff Mentch, Jerry Ng, Matthew Beveridge, and Iddo Drori.
\newblock Image2reverb: Cross-model reverb impulse response synthesis.
\newblock In \emph{IEEE/CVF International Conference on Computer Vision
  (ICCV)}, October 2021.

\bibitem[Du et~al.(2021)Du, Collins, Tenenbaum, and Sitzmann]{du2021gem}
Yilun Du, M.~Katherine Collins, B.~Joshua Tenenbaum, and Vincent Sitzmann.
\newblock Learning signal-agnostic manifolds of neural fields.
\newblock In \emph{Advances in Neural Information Processing Systems}, 2021.

\bibitem[Engel et~al.(2019)Engel, Agrawal, Chen, Gulrajani, Donahue, and
  Roberts]{engel2019gansynth}
Jesse Engel, Kumar~Krishna Agrawal, Shuo Chen, Ishaan Gulrajani, Chris Donahue,
  and Adam Roberts.
\newblock Gansynth: Adversarial neural audio synthesis.
\newblock \emph{arXiv preprint arXiv:1902.08710}, 2019.

\bibitem[Paasonen et~al.(2017)Paasonen, Karapetyan, Plogsties, and
  Pulkki]{paasonen2017proximity}
Juhani Paasonen, Aleksandr Karapetyan, Jan Plogsties, and Ville Pulkki.
\newblock Proximity of surfaces—acoustic and perceptual effects.
\newblock \emph{Journal of the Audio Engineering Society}, 65\penalty0
  (12):\penalty0 997--1004, 2017.

\bibitem[Straub et~al.(2019)Straub, Whelan, Ma, Chen, Wijmans, Green, Engel,
  Mur-Artal, Ren, Verma, et~al.]{straub2019replica}
Julian Straub, Thomas Whelan, Lingni Ma, Yufan Chen, Erik Wijmans, Simon Green,
  Jakob~J Engel, Raul Mur-Artal, Carl Ren, Shobhit Verma, et~al.
\newblock The replica dataset: A digital replica of indoor spaces.
\newblock \emph{arXiv preprint arXiv:1906.05797}, 2019.

\bibitem[Koyama et~al.(2021)Koyama, Nishida, Kimura, Abe, Ueno, and
  Brunnstr{\"o}m]{koyama2021meshrir}
Shoichi Koyama, Tomoya Nishida, Keisuke Kimura, Takumi Abe, Natsuki Ueno, and
  Jesper Brunnstr{\"o}m.
\newblock Meshrir: A dataset of room impulse responses on meshed grid points
  for evaluating sound field analysis and synthesis methods.
\newblock In \emph{2021 IEEE Workshop on Applications of Signal Processing to
  Audio and Acoustics (WASPAA)}, pages 1--5. IEEE, 2021.

\bibitem[Savioja et~al.(1999)Savioja, Huopaniemi, Lokki, and
  V{\"a}{\"a}n{\"a}nen]{savioja1999creating}
Lauri Savioja, Jyri Huopaniemi, Tapio Lokki, and Ritta V{\"a}{\"a}n{\"a}nen.
\newblock Creating interactive virtual acoustic environments.
\newblock \emph{Journal of the Audio Engineering Society}, 47\penalty0
  (9):\penalty0 675--705, 1999.

\bibitem[Raghuvanshi et~al.(2010)Raghuvanshi, Snyder, Mehra, Lin, and
  Govindaraju]{raghuvanshi2010precomputed}
Nikunj Raghuvanshi, John Snyder, Ravish Mehra, Ming Lin, and Naga Govindaraju.
\newblock Precomputed wave simulation for real-time sound propagation of
  dynamic sources in complex scenes.
\newblock In \emph{ACM SIGGRAPH 2010 papers}, pages 1--11. 2010.

\bibitem[P{\"o}rschmann et~al.(2020)P{\"o}rschmann, Arend, Bau, and
  L{\"u}beck]{porschmann2020comparison}
Christoph P{\"o}rschmann, Johannes~M Arend, David Bau, and Tim L{\"u}beck.
\newblock Comparison of spherical harmonics and nearest-neighbor based
  interpolation of head-related transfer functions.
\newblock In \emph{Audio Engineering Society Conference: 2020 AES International
  Conference on Audio for Virtual and Augmented Reality}. Audio Engineering
  Society, 2020.

\bibitem[D{\'e}fossez et~al.(2018)D{\'e}fossez, Zeghidour, Usunier, Bottou, and
  Bach]{defossez2018sing}
Alexandre D{\'e}fossez, Neil Zeghidour, Nicolas Usunier, L{\'e}on Bottou, and
  Francis Bach.
\newblock Sing: Symbol-to-instrument neural generator.
\newblock \emph{arXiv preprint arXiv:1810.09785}, 2018.

\bibitem[Kolarik et~al.(2016)Kolarik, Moore, Zahorik, Cirstea, and
  Pardhan]{kolarik2016auditory}
Andrew~J Kolarik, Brian~CJ Moore, Pavel Zahorik, Silvia Cirstea, and Shahina
  Pardhan.
\newblock Auditory distance perception in humans: a review of cues,
  development, neuronal bases, and effects of sensory loss.
\newblock \emph{Attention, Perception, \& Psychophysics}, 78\penalty0
  (2):\penalty0 373--395, 2016.

\end{thebibliography}
